\documentclass[12pt,a4paper]{article}
\usepackage[cmex10]{amsmath}
\usepackage{siunitx}
\usepackage{graphicx}
\usepackage{array}
\usepackage{color}
\usepackage{fixltx2e}
\usepackage{float}

\title{$Q$-enhanced racetrack microresonators}

\author{ P. Chamorro-Posada\thanks{Departmento de Teor\'{\i}a de la Se\~nal y Comunicaciones e Ingenier\'{\i}a Telem\'atica, Universidad de Valladolid, ETSI Telecomunicaci\'on, Paseo Bel\'en 15, E-47011 Valladolid, Spain. }}

\date{\today}
\begin{document}

\maketitle

\begin{abstract}

A $Q$-enhancement strategy for racetrack microresonators is put forward.  The design is based on the modification of the resonator geometry in order to mitigate the two main sources of radiation loss in the presence of curved waveguides: the discontinuities at the junctions between straight waveguides and the bent sections, and the continuous loss at the curved waveguide sectors.   At the same time, the modifications of the geometry do not affect the versatility of coupling of racetrack resonators in integrated optical circuits, which is their main advantage over ring microresonators.  The proposal is applied to the design of high-$Q$ racetrack resonators for the silicon nitride CMOS-compatible platform having bent radii amenable for large-scale photonic integration.  Numerical calculations show over $100\%$ improvement of the $Q$ factor in Si\textsubscript{3}N\textsubscript{4}/SiO\textsubscript{2} resonators. 

\end{abstract}

\section{Introduction}

Racetrack and ring microresonators are highly versatile elements within integrated photonics with applications as add-drop multiplexers \cite{little1}, optical filters \cite{little2}, optical switches \cite{almeida}, sensors \cite{vos}, modulators \cite{xu}, or in slow light systems based on coupled resonator waveguides \cite{scheuer}.  These microresonators can be implemented in a variety of optical integration platforms: silicon on insulator (SOI)\cite{bogaerts}, silica on silicon \cite{ou}, polymer \cite{rabiei}, Si\textsubscript{3}N\textsubscript{4}/SiO\textsubscript{2} \cite{tien}, GaAs/AlGaAs \cite{van}, or InP/InGaAsP \cite{grover}. 

A  key parameter characterizing the performance of a resonator is the quality factor $Q$ \cite{bogaerts}.  In the time domain, the $Q$-factor of a resonant mode is determined by the $1/e$ decay time of the electromagnetic energy stored in that mode.  In the frequency domain, it is equivalently characterized by the sharpness of the resonance relative to its central frequency.  Its value depends on the total loss, including both the effects due to coupling to an external circuit and those associated to the propagation in the ring. In the absence of external coupling, the intrinsic $Q$-factor is called the unloaded $Q$ of the resonator.  The coupling of the microresonator to an external waveguide modifies the resonator $Q$ to its loaded value $Q_L$.  Although in certain applications, such as optical delay lines, a low value of $Q_L$ is usually desirable to increase the transmission bandwidth \cite{khurgin}, any reduction of the intrinsic $Q$ in this cases results in performance degradation and high values of the unloaded $Q$, limited by the radiation and/or transmission losses, are most typically advantageous.  

There are various contributions to the round-trip loss in an unloaded resonator: the intrinsic propagation loss of the waveguide material, the effect of the roughness of the waveguide walls or the effects due to the bending in the curved waveguide sections.  In high contrast integrated optics platforms, the effect of bending can be negligible even for very small radii of curvature.  Such is the case in SOI photonic circuits.  In this platform, on the other hand, the intrinsic propagation losses are relatively high.  The converse situation is found, for instance, in Si\textsubscript{3}N\textsubscript{4}/SiO\textsubscript{2} integrated circuits, where intrinsic losses are very small, but the reduced waveguide core/cladding refractive index contrast relative to that of SOI can result in higher values of the radiation loss due to waveguide curvature.

Ring microresonators offer higher values of $Q$ when compared with racetrack microresonators.  For a fixed FSR, the bending radius is larger in the ring configuration, which is also free from the discontinuities at the junctions between the straight and curved sections.  Nevertheless, the straight sections in the racetrack geometry permit a more accurate control of the coupling to external waveguides that is limited by the fabrication tolerances.  This often makes the racetrack arrangement the preferred option \cite{haus}.   

A pulley type configuration, with the coupling bus waveguide surrounding the resonator, has been shown to reduce the radiation loss and to provide very high $Q$ factors when used in microdisk \cite{hosseini,hu} and microring \cite{cai} resonators, even though improvements have been more elusive in the case of racetrack microresonators \cite{cai}. 

In the pulley geometry, the bus waveguide is a design element both of the resonator and the coupling system, resulting a largely restrictive configuration.  In this work, we exploit the radiation quenching properties of the external slab in pulley resonators but  with independence to the coupling of the resonator.  This is supplemented with the lateral offset \cite{kitoh} technique to reduce the loss due to the transition from the straight to the bent transmission section, which had been previously addressed for GaAs-AlGaAs resonators \cite{van2}.  The resulting geometry is shown to provide large improvements in the $Q$ factors of racetrack microresonators.  This is particularly interesting when the radiation loss can dominate the intrinsic propagation loss, as it is the case in silicon nitride photonic integrated circuits.  At the same time, the versatility of racetrack microresonators as key design elements of integrated optical circuits is kept intact, since the geometry modifications affect at the resonator itself and not the coupling properties to the optical circuit.

\section{The silicon nitride platform: device fabrication considerations}

The photonic integration platforms that exploit the broadly established complementary metal-oxide-semiconductor (CMOS) infrastructure are particularly appealing.  Among them, the silicon-on-insulator (SOI) \cite{bogaerts} is the most developed one, and has footed the development of an incipient silicon photonics industry.  Large scale photonic integration is facilitated by the large index contrast of SOI waveguides.  Nevertheless, other limiting factors, such as two-photon absorption, affect nonlinear photonic applications.  Silicon nitride is an emerging CMOS-compatible alternative to SOI photonics.  Even though the refractive index contrast is smaller than that of SOI, it offers reduced intrinsic linear \cite{bauters} and nonlinear losses \cite{moss}.  Very small linear losses are very interesting, for instance, for quantum applications operating at the single-photon level \cite{xiong}.  Low nonlinear losses, on the other hand, are very appealing for nonlinear photonics applications \cite{moss}.  Further applications of the silicon nitride platform include coherence tomography \cite{yurtsever} or lab-on-a-chip devices \cite{cai2,ymeti}.

\begin{figure}[H]
\centering
\includegraphics[width=2.5in]{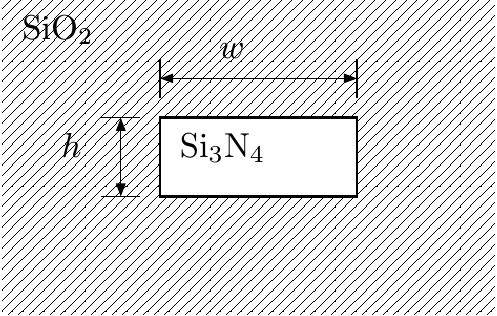}
\caption{Channel waveguide geometry used for the numerical calculations.}
\label{fig::guia}
\end{figure}

 We will assume a typical Si\textsubscript{3}N\textsubscript{4}/SiO\textsubscript{2} channel waveguide geometry  as shown in Fig. \ref{fig::guia}.   In order to obtain high optical quality deposited films, the silicon nitride layer height $h$ is limited to values of $h<\SI{400}{\nano\metre}$ due to film stress.  Catastrophic cracking occurs at thicker layers, severely limiting device performance \cite{luke}.  Extremely thin silicon nitride films \cite{tien2} avoid stress issues.  More sophisticated fabrication processes, for instance, with the introduction of mechanical trenches for isolating photonic devices from propagating cracks \cite{luke}, permit to grow thicker ($h>\SI{400}{\nano\metre}$) Si\textsubscript{3}N\textsubscript{4} layers with better confined optical modes and smaller radiation losses. 

 Ultra-high $Q$ silicon nitride ring microresonators with very thin Si\textsubscript{3}N\textsubscript{4} layers have highly delocalized modes and require bend radii in the millimeter range  \cite{tien2}.   Even for thick Si\textsubscript{3}N\textsubscript{4} films  with improved mode confinement, ring radii are still over one hundred microns \cite{luke}.  This contrasts with high-$Q$ silicon ring microresonators that can have radii comparable to the optical wavelength in vacuum \cite{xu2}.  Large rings not only hinder large scale integration, but the corresponding reduction in the free spectral range can also impose a severe limitation for certain applications.  As discussed below, one of the radiation quenching geometry modifications can also be applied to ring microresonators, even though this work focuses specifically on the racetrack geometry. 

In the design of complex integrated photonic circuits, racetrack microresonators offer superior versatility over ring microresonators due to the better control of the coupling coefficients to the external optical circuitry, but they suffer from increased radiation losses due to the transitions of straight to bent waveguide sections.  In this work, we will seek practical values of the bent radii of the racetrack resonators compatible with large scale integration in the silicon nitride platform by the implementation of mitigation measures for the sources of radiation loss.

We will assume an intermediate channel waveguide geometry (similar, for instance, to that of \cite{gondarenko}), with relatively high mode confinement within the film stress limits.  The parameters assumed are $w=$\SI{1}{\micro\metre} and $h=$\SI{300}{\nano\metre}.   The refractive indices of silica and silicon nitride have been taken as $1.4501$ and $1.9792$, respectively. With these parameters, the devices support both quasi-TE and quasi-TM polarized modes, but coupling between the two polarizations is expected to be negligible \cite{luke}.  In the analysis, we will focus on the lowest order quasi-TE polarized mode.

In the fabrication process, there is a deviation between the actual waveguide width and the waveguide within the mask layout; the so-called underetch. This deviation is due to mask erosion and the etching processes, but its effect can be handled automatically in the mask layout by the rendering software. Besides the underetch, there is also a variation in the waveguide width across the wafer due to the lithography process. The waveguide-width tolerance across the wafer can be typically of $\SI{100}{\nano\metre}$.

\section{Proposed scheme}

There exist two main and independent radiation loss mechanisms associated with the propagation in curved waveguides \cite{lewin,heiblum,marcuse}.  The first one is the coupling mismatch existing between the optical mode field in the straight waveguide section and that of a curved waveguide with constant curvature.  The second one is the continuous radiation loss of the modal field in the curved transmission sections.

\begin{figure}[H]
\centering
\begin{tabular}{c}
{\large (a)}\\
\includegraphics[width=3in]{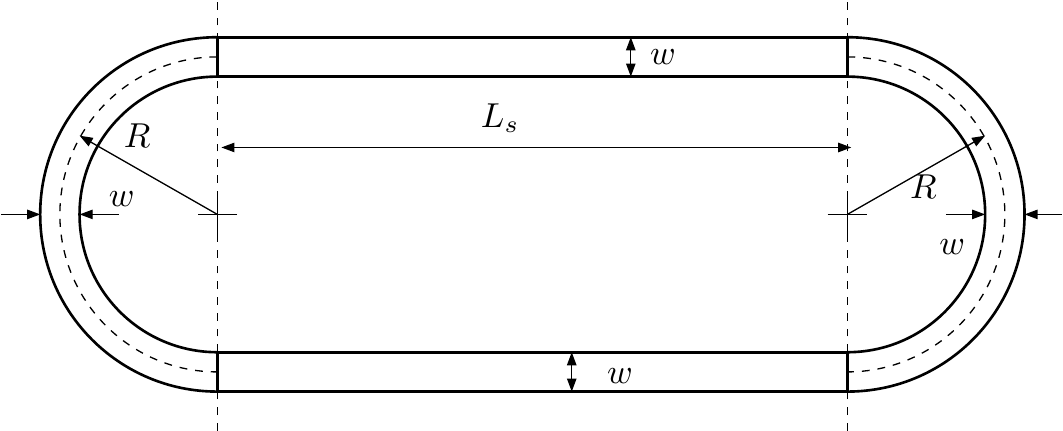}\\
{\large (b)}\\
\includegraphics[width=3in]{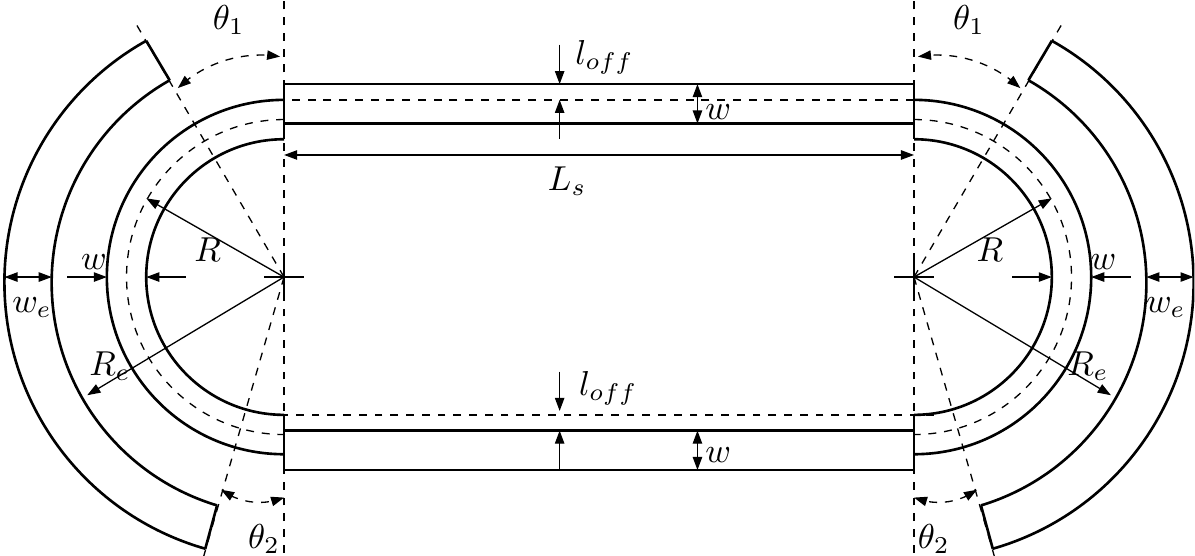}
\end{tabular}
\caption{Conventional (a) and $Q$-enhanced (b) racetrack resonator geometries.}
\label{fig::geometria}
\end{figure}

Since these two sources of radiation loss have clearly distinct physical origin, they are addressed independently in the proposed scheme and any possible correlation effect in the geometry modifications is neglected at the design stage.  The losses at the waveguide transitions are dealt with using the lateral offset technique \cite{kitoh}.  For the radiation of the bent sections, the intrinsic radiation quenching mechanisms of pulley resonators \cite{cai} is adopted.   With an adequate design, the waveguide around the resonator in a pulley configuration has the effect of reducing the radiation loss when the modes in the inner and outer coupled structures are phase mismatched.  This is similar to the result of phase mismatch in asymmetric couplers \cite{marcusec}, where the coupler supermodes do not have the typical even and odd distributions of symmetric couplers, but are mainly localized at one of the guides and evanescent at the other.  For a bent guiding structure the presence of a nonsynchronous parallel curved slab has the effect of reducing the radiation loss \cite{cai}.     

\begin{figure}[H]
\centering
\begin{tabular}{cc}
{\large (a)}&{\large (b)}\\
\includegraphics[width=2.5in]{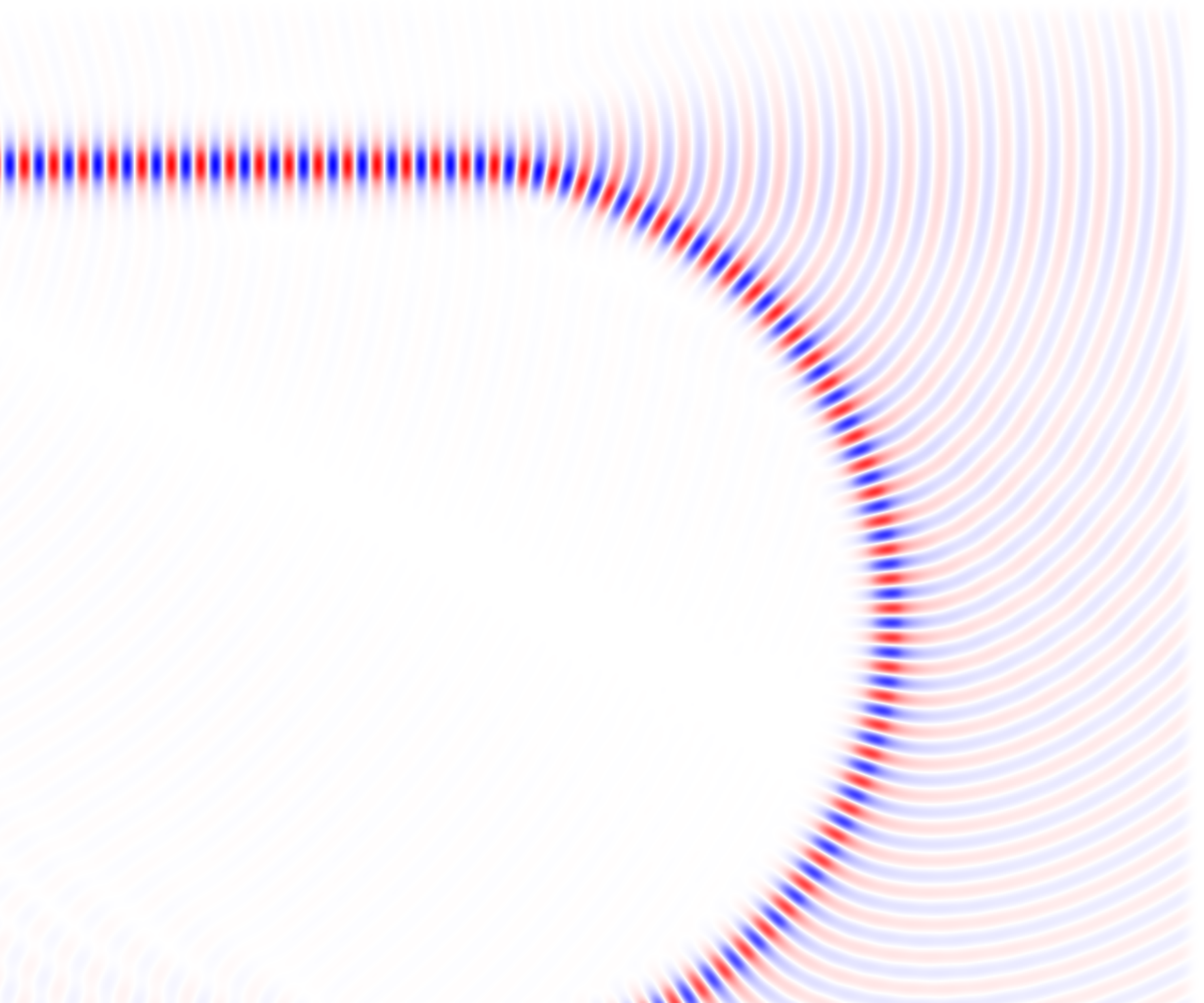}&\includegraphics[width=2.5in]{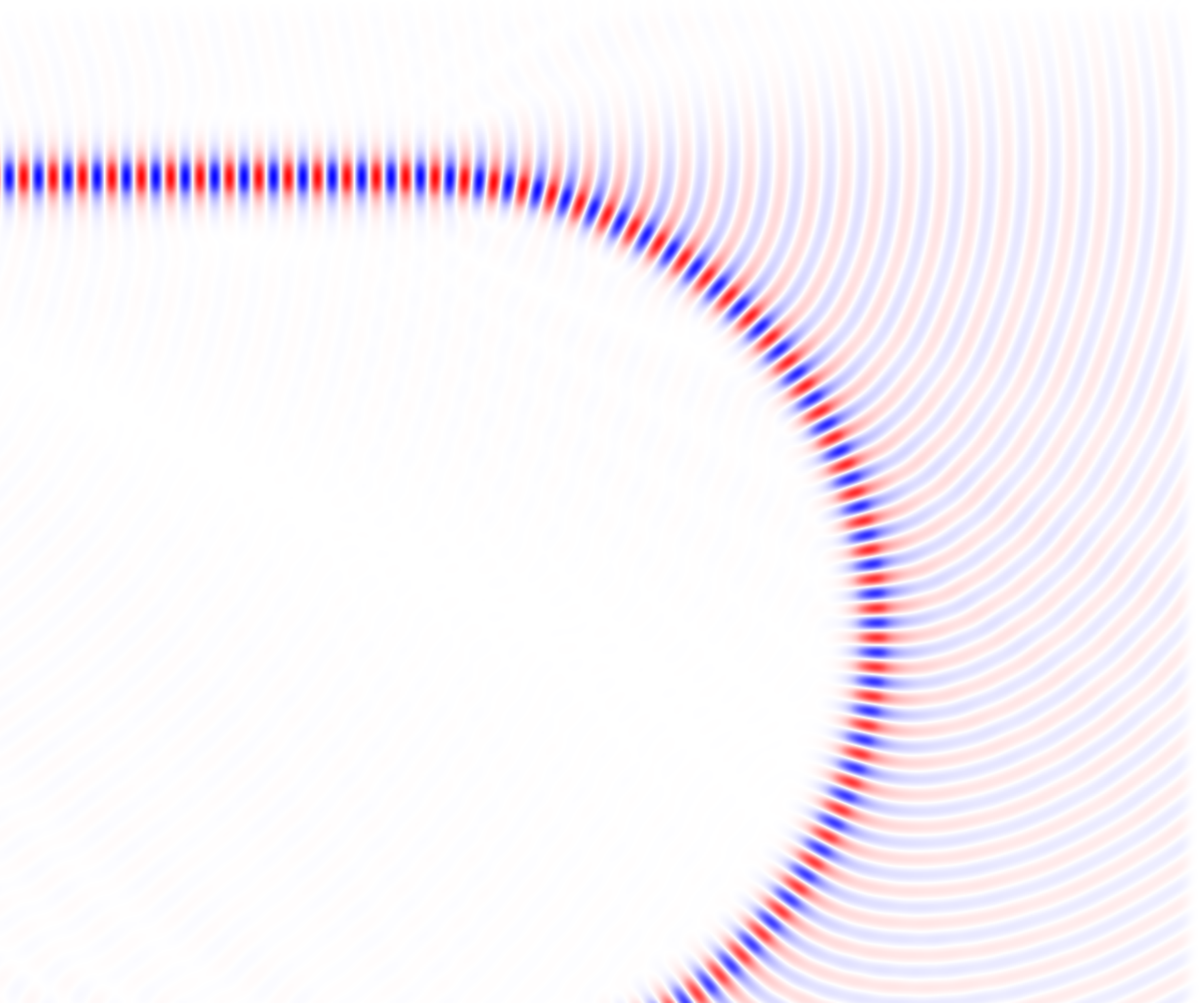}\\
{\large (c)}&{\large (d)}\\
\includegraphics[width=2.5in]{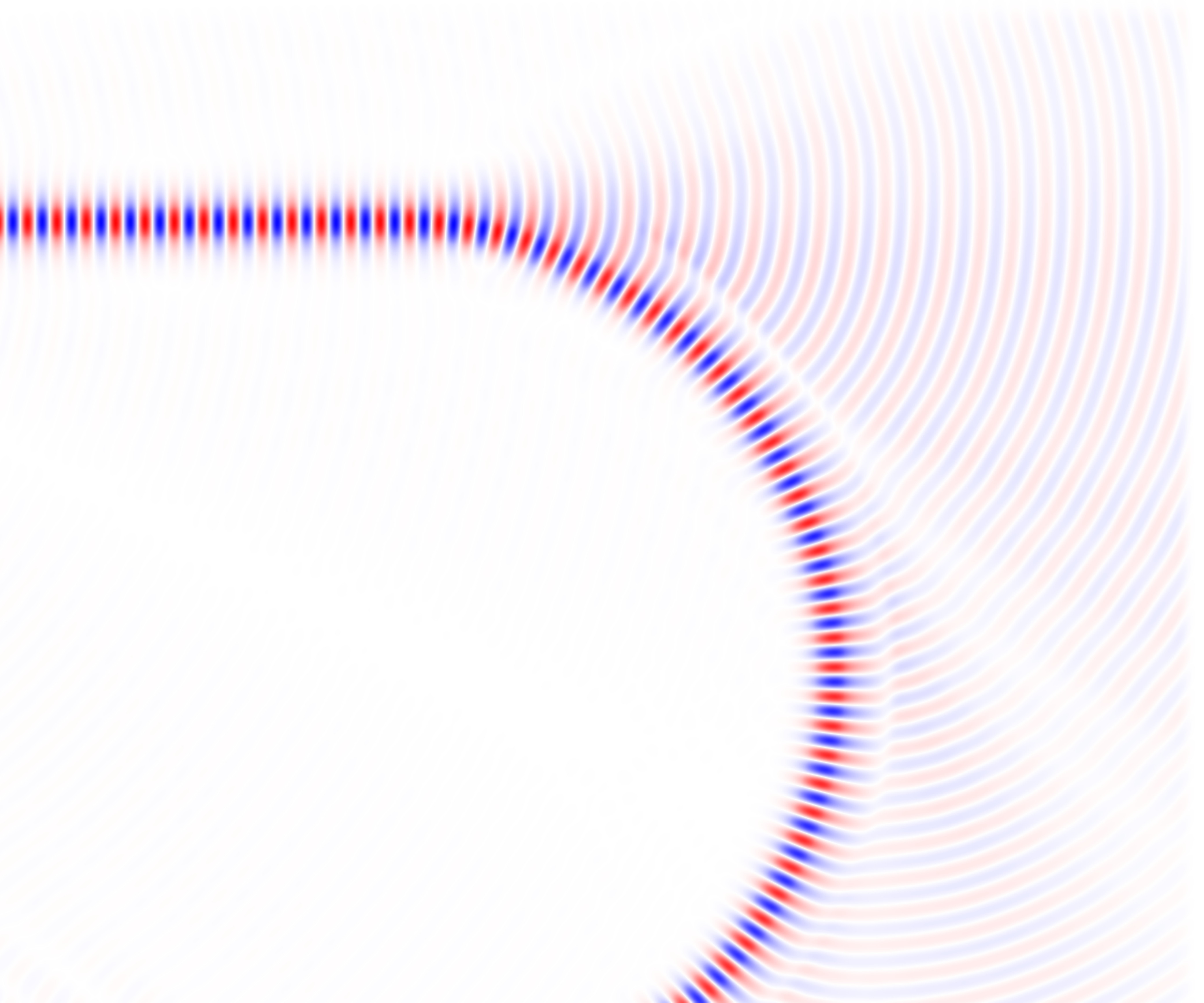}&\includegraphics[width=2.5in]{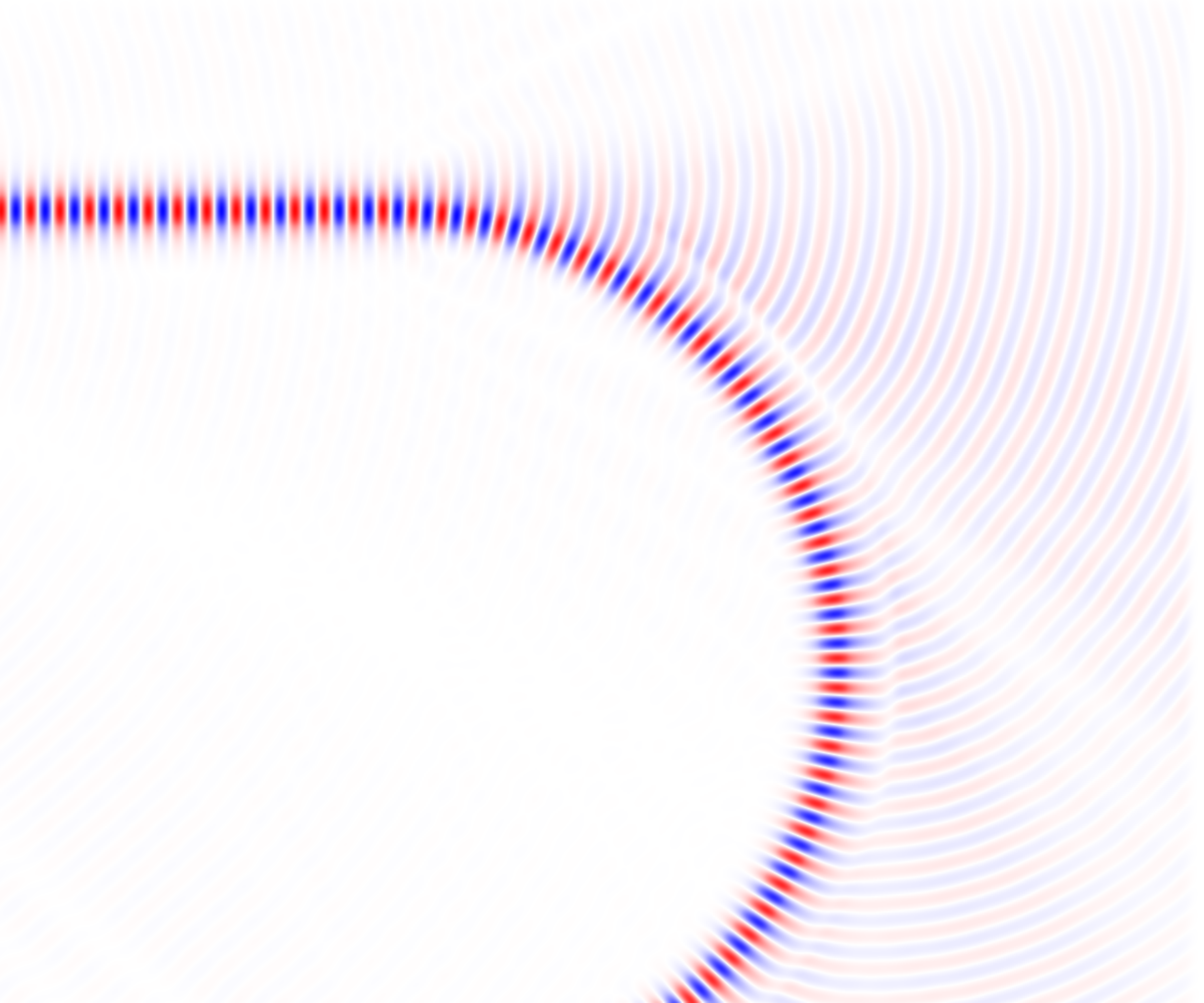}
\end{tabular}
\caption{Radiation patterns under continuous-wave excitation close to the straight-bent transition for (a) the unmodified geometry, (b) when the lateral offset is used, (c) when the external annulus is employed and (b) when both lateral offset and external annulus are introduced.}
\label{fig::efeto}
\end{figure}

The conventional geometry of a racetrack microresonator is shown in Fig. \ref{fig::geometria} (a).  This geometry is fully defined by the waveguide width $w$, the radius of the bent sections $R$ and the length of the straight sections $L_s$.  It can be compared with the proposed scheme in Fig. \ref{fig::geometria} (b).  The modifications include a lateral shift $l_{off}$ of the straight waveguides and the existence of radiation quenching curved exterior sectors of width $w_e$ and inner radius of curvature $R_e$.  The nonsynchronous condition of the curved coupler will require that either $w_e<w$ or $w_e>w$.  The angular extent of the exterior sectors is limited by the angular parameters $\theta_l$, $l=1,2$, that are intended to be adjusted to permit the coupling of the resonator to other structures either from one side or both sides.  Large values of $\theta_l$ are suboptimal in relation with the maximal reduction in the radiation loss.  This effect is studied in the following section.       

The effect of the geometry modifications on the radiation patterns near the straight-to-bent waveguide junction can be visualized in Fig. \ref{fig::efeto}.  For these results, the continuous-wave excitation vacuum wavelength is $\lambda_0=\SI{1.55}{\micro\metre}$ and the bend radius is $R=\SI{15}{\micro\metre}$.  Fig. \ref{fig::efeto} (a) displays the main features of this radiation, with one component beamed along the axis of the straight incident waveguide and a second component continuously shed in the bent section.  The introduction of the lateral offset of $l_{off}=\SI{0.9}{\micro\metre}$ permits to match the mode field in the straight waveguide to that of the curved section, with a main lobe displaced towards the waveguide edge.  This drastically reduces the beam-like radiation component in Fig. \ref{fig::efeto} (b).  Fig. \ref{fig::efeto} (c) shows the effect of the introduction of an exterior ring with a radius $R_e=\SI{17}{\micro\metre}$ and a width $w_e=\SI{400}{\nano\metre}$. In this case, it is the continuous radiation contribution due to the waveguide curvature the one that is largely reduced.  Finally, Fig. \ref{fig::efeto} (d) shows the combined effect of the two measures.

\begin{figure}[h]
\centering
\begin{tabular}{cc}
{\large (a)}&{\large (b)}\\
\includegraphics[width=1.5in]{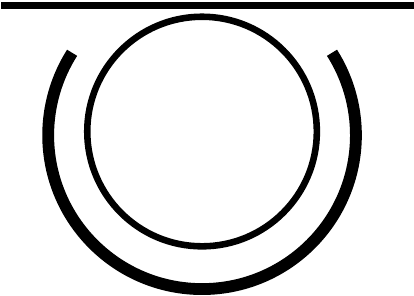}&
\includegraphics[width=1.5in]{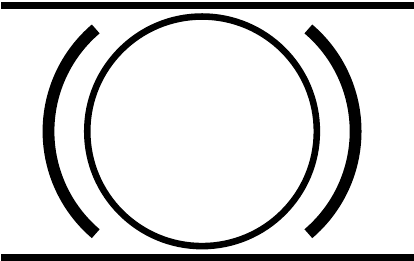}
\end{tabular}
\caption{Radiation reduction by an external slab in microring resonators.}
\label{fig::ring}
\end{figure}

The reduction of the radiation loss and the increase of the $Q$-factor by using an external slab can also be employed in microring resonators, as shown in Figures \ref{fig::ring} (a) and (b) for a side coupled and add-drop microring, respectively.

\section{Modeling}

The 3D buried channel waveguide geometry shown in Fig. \ref{fig::guia} is reduced to a 2D problem using the effective index method \cite{tamir}.  We focus on the quasi-TE mode field polarization, with the electric field predominantly polarized along the wider dimension in Fig. \ref{fig::guia}.  First, the transverse problem is solved in the vertical direction, where a single-mode symmetric slab waveguide with TE polarization and core width $h$ describes the propagation in the central region of the 2D problem.  The final effective index model is then obtained looking again at the transverse waveguide geometry, but now along the horizontal direction with the core described by the effective index of the mode obtained in the former analysis. Since the electric field is now directed normal to the waveguide boundaries, the evolution in the resulting model corresponds to that of 2D TM field. 

\begin{figure}[h]
\centering
\includegraphics[width=3in]{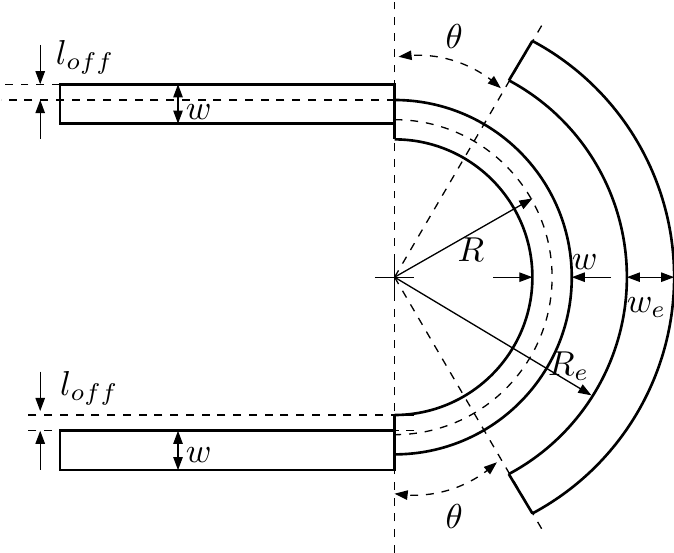}
\caption{Geometry employed in the parametric search for the optimization of the structure.}
\label{fig::opt}
\end{figure}

Solutions of the 2D equivalent model are obtained using the finite-difference time-domain (FDTD) method as implemented in the MEEP software package \cite{meep}. FDTD simulations are highly accurate and versatile, easily permitting to study the propagation effects in curved waveguide sections.  On the other hand, the computational requirements are far larger than those of other numerical techniques.  The existence of four parameters in the optimization domain ($l_{off}$, $R_e$, $w_e$, $\theta$) for the device geometry shown in Fig. \ref{fig::opt}  together with the high computational load of FDTD method demands a judicious strategy that allows for a systematic, affordable, optimization of the microresonator geometries.

The proposal itself is the result of the combination of two geometry modifications that independently address two distinct physical effects that produce radiation losses, even though both are are associated to the presence of curved waveguide sections.  The effect of the discontinuity from straight to bent sections in the racetrack that is mitigated with the lateral offset technique would be absent in a ring microresonator.  At the same time, the radiation continuously shed at the curved propagation sections can be affected by the coupling condition at the discontinuity, but it is expected to be mostly dependent on the modal field properties at the curved waveguides.

Even though the simultaneous presence of the two loss mitigation measures may result in some mutual interdependence, the device optimization is performed under the assumption that the loss effects can be treated independently.  The optimal lateral offset $l_{off}$ is calculated in the absence of the pulley-type waveguide.  Also, the dependence of the radiation quenching properties with the pulley exterior ring parameters $R_e$ and $w_e$ is calculated with $l_{off}=0$.  Besides, this optimization is performed with fixed  $\theta_1=\theta_2=\theta$.  The value of  $\theta$ is physically limited by the presence of the access waveguides in the device fabrication.  A reasonable value of $\theta=\SI{30}{\degree}$ is used in most of the calculations.  Nevertheless, the effect of the value of $\theta$ are also studied independently for certain parameter values.  In a latter analysis, the combined effects in the resonator Q-factor obtained from the independently optimized parameters are analyzed for several configurations.  The optimization strategy, though limited by the intense computational load of FDTD calculations, is shown to produce valuable designs with large improvements in the resonator $Q$s.

The proposal is evaluated at two representative values of the bend radius: $R=$\SI{15}{\micro\metre} and $R=$\SI{25}{\micro\metre}.  It is in this regime, with significative radiation losses, where the proposed geometry modifications are relevant and correspond to a meaningful reduction on the bend radii over conventional implementations.  Furthermore, other sources of propagation loss are negligible when compared to the radiation losses and can be safely neglected.  In any case, other loss mechanisms would produce a fixed contribution to the reduction of the intrinsic resonator $Q$ which would equally affect all the cases and, therefore, are irrelevant for our optimization. 

The geometry employed in the optimization stage is shown in Fig. \ref{fig::opt}.  It corresponds to the propagation along half a resonator round-trip.  A narrow Gaussian optical pulse is injected in the top straight waveguide section and the time-domain output field is measured in the bottom straight waveguide.  Fourier transforming the temporal response permits to determine the propagation loss in each configuration in a broad spectral range.  In the analysis, the power round-trip loss $A$  is calculated for each case from the FDTD simulations and the percentual improvement over the case without geometry modifications $A_0$ is employed for performance evaluation purposes.  Direct percentual improvements would be very difficult to interpret, since the absolute radiation loss varies widely between the two radii considered.  A fair comparison can be obtained if the power lost due to radiation in one propagation round-trip of duration $t_R$ is related with the cavity life-time $\tau_0$ as
\begin{equation}
A=\exp\left(-2\dfrac{t_R}{\tau_0}\right).
\end{equation}      
In turn, $\tau_0$ is directly related with the cavity intrinsic $Q_0$ as \cite{haus}
\begin{equation}
Q_0=\dfrac{\omega_0\tau_0}{2},
\end{equation}
 where $\omega_0=2\pi\nu_0$ and $\nu_0$ is the frequency of the optical carrier. The relative round trip loss improvement $IF$ used in this work is defined as 
\begin{equation}
IF \equiv \dfrac{\log A}{\log A'}=\dfrac{\tau_0'}{\tau_0}=\dfrac{Q_0'}{Q_0},\label{eq::if}
\end{equation}
where results with prime correspond to the transmission in the presence of radiation mitigation geometry modifications.  This way, also, the improvement factors obtained from the radiation reduction strategies are quantified by their direct impact on the cavity intrinsic $Q_0$.

\begin{figure}[H]
\centering
\begin{tabular}{c}
{\large (a)}\\
\includegraphics[width=3in]{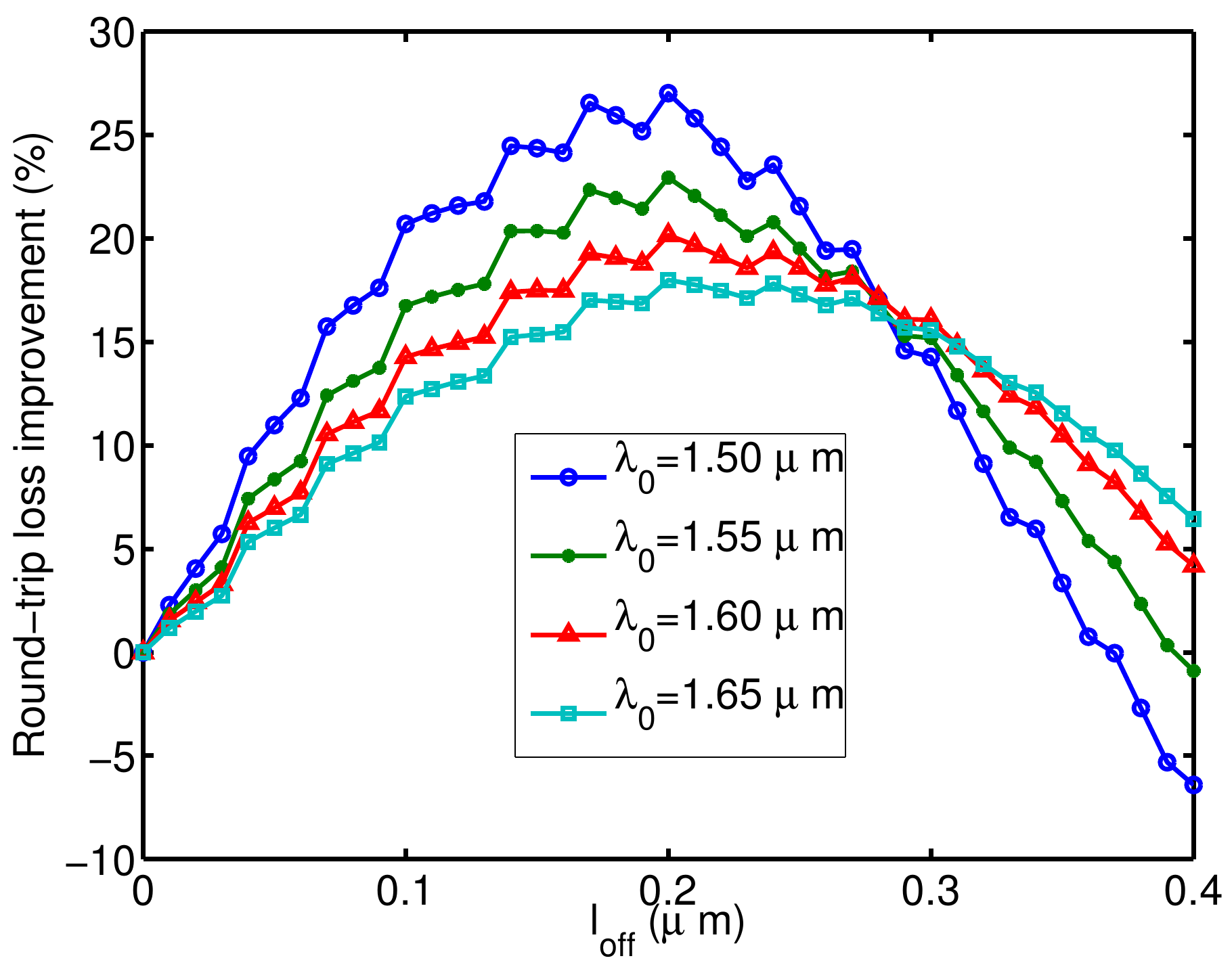}\\
{\large (b)}\\
\includegraphics[width=3in]{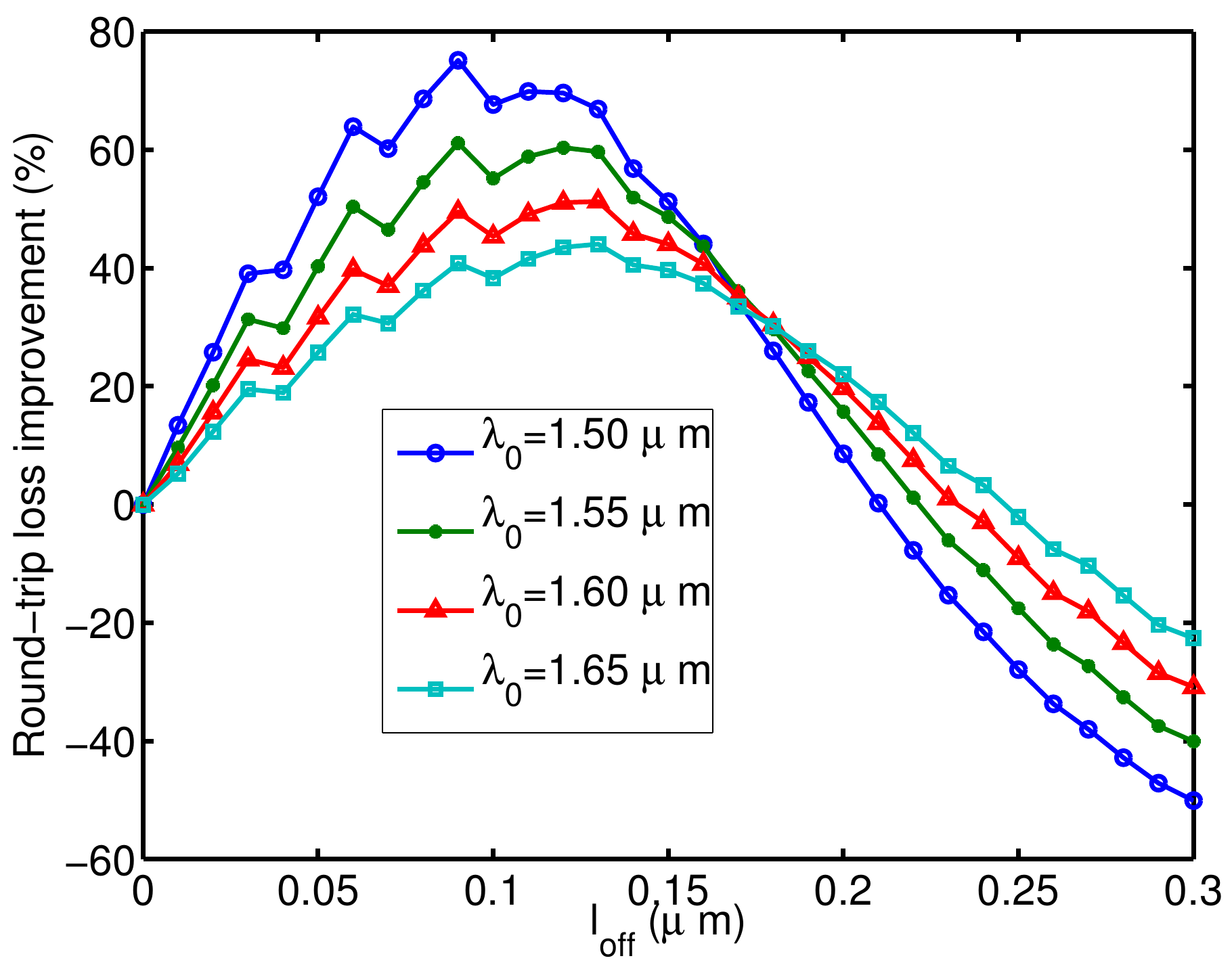}\\
\end{tabular}
\caption{Round-trip improvement factor for the geometry displayed in Fig. \ref{fig::opt} as a function of the lateral offset $l_{off}$ for $R=$\SI{15}{\micro\metre} (a) and $R=$\SI{25}{\micro\metre} (b).  In both cases, $w_e=0$.  Results are plotted for four values of the vacuum wavelength. }
\label{fig::offset}
\end{figure}

\section{Results}

\subsection{Geometry optimization}

\subsubsection{Lateral offset}

The calculated round-trip improvement factors, as defined in Eq. \eqref{eq::if}, as a function of the lateral offset $l_{off}$ (see Fig. \ref{fig::opt})  are displayed in Fig. \ref{fig::offset} for $R=$\SI{15}{\micro\metre} (a) and $R=$\SI{25}{\micro\metre} (b).  Results are plotted for four vacuum wavelength values covering the whole band between \SIrange{1.50}{1.65}{\micro\metre}.  As a whole, the results show a monotonic increase of the peak improvement factor with $R/\lambda_0$.  Therefore, the improvement factors attainable for $R=$\SI{25}{\micro\metre} radius are significantly larger than those for $R=$\SI{15}{\micro\metre} in this band.  Nevertheless, the sensitivity to the optimal value of $l_{off}$ also increases with $R/\lambda_0$.  Whereas for  $R=$\SI{25}{\micro\metre} fabrication tolerances would have a negligible on the performance of fabricated devices, they would start to be an issue for the shorter wavelengths at  $R=$\SI{25}{\micro\metre}.

\subsubsection{Pulley ring sector}

\begin{figure}[h]
\centering
\begin{tabular}{cc}
{\large (a)}&{\large (b)}\\
\includegraphics[width=2.5in]{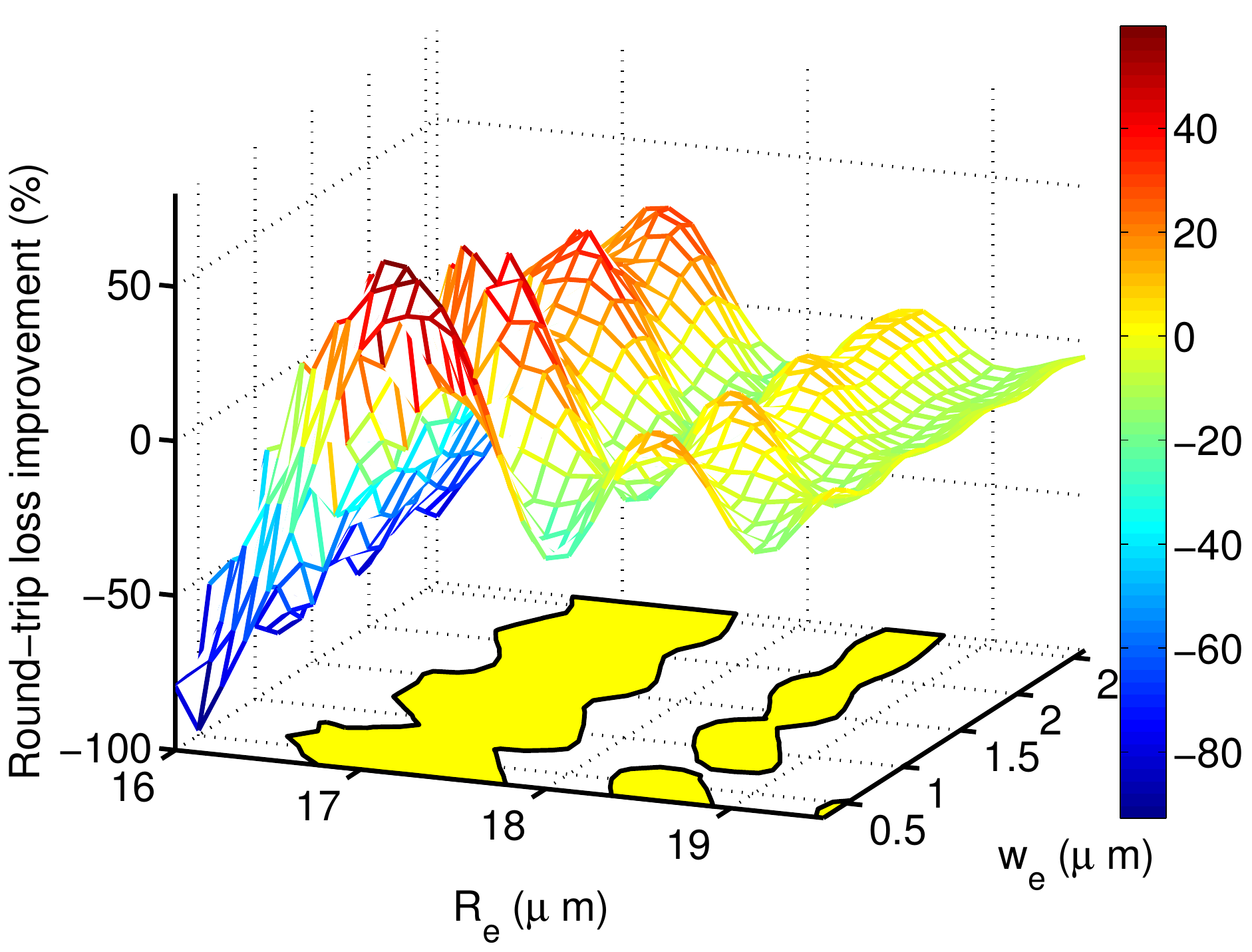}&\includegraphics[width=2.5in]{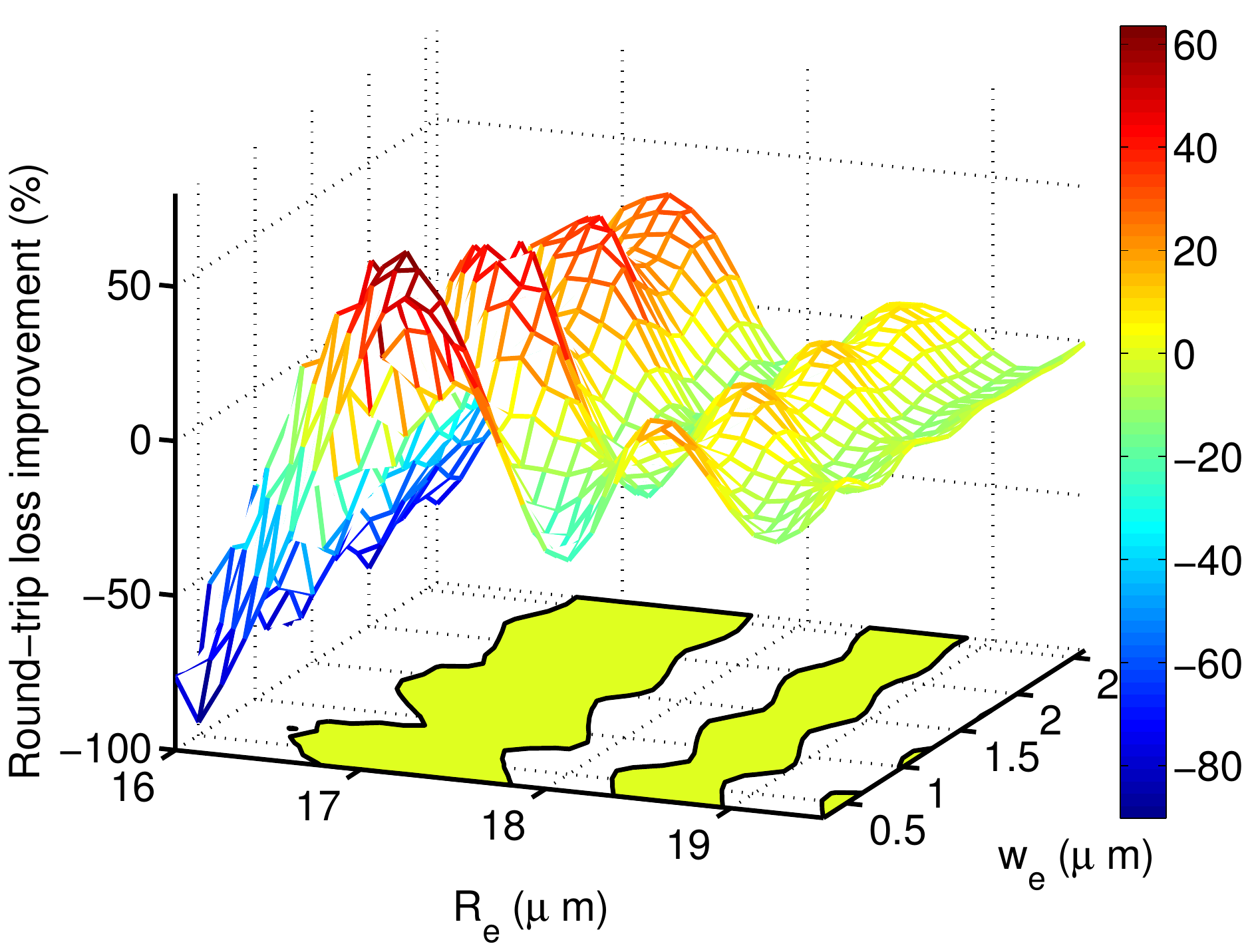}\\
{\large (c)}&{\large (d)}\\
\includegraphics[width=2.5in]{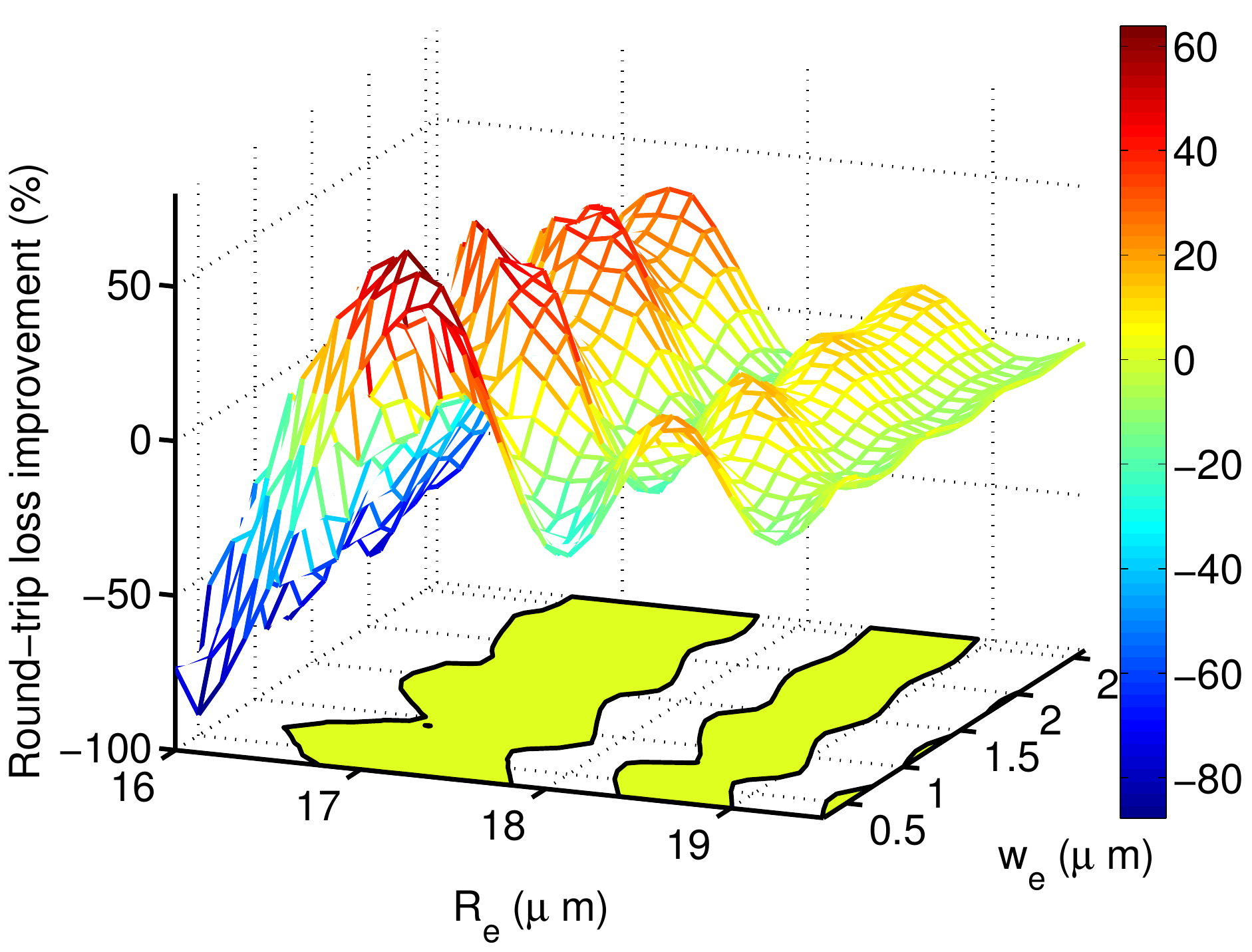}&\includegraphics[width=2.5in]{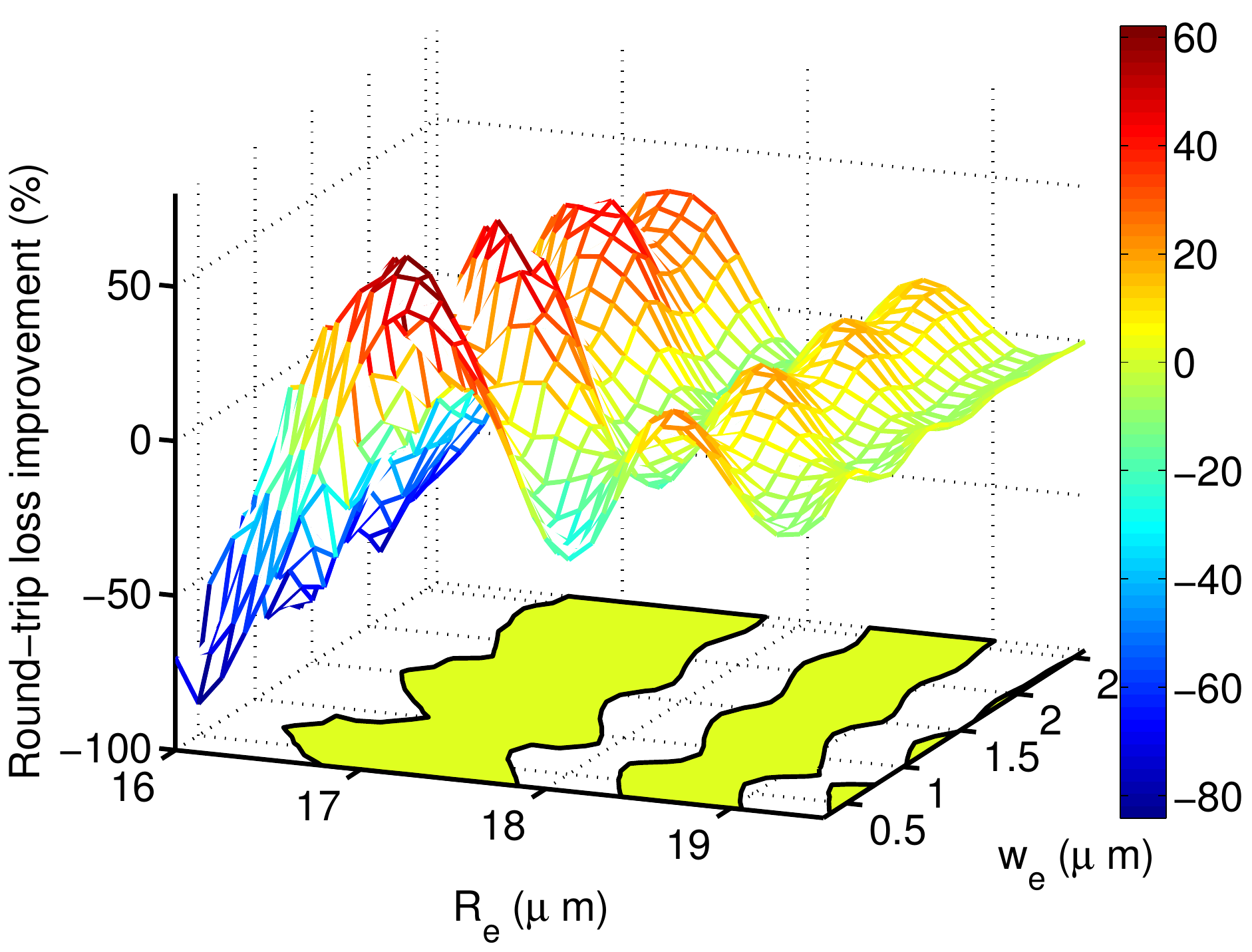}
\end{tabular}
\caption{Round-trip loss improvement factors as a function of the exterior annuli parameters $R_e$ and $w_e$ for $R=$\SI{15}{\micro\metre}  and $\theta=\SI{30}{\degree}$.  Four vacuum wavelength values are considered: (a) $\lambda_0=$\SI{1.50}{\micro\metre}, (b) $\lambda_0=$\SI{1.55}{\micro\metre}, (c) $\lambda_0=$\SI{1.60}{\micro\metre}  and (d) $\lambda_0$=\SI{1.65}{\micro\metre}.  }
\label{fig::exteriorR15}
\end{figure}

Fig. \ref{fig::exteriorR15} depicts the loss improvement factors as a function of the parameters defining the exterior annulus ($R_e$ and $w_e$) for $R=$\SI{15}{\micro\metre} and a fixed value of $\theta=\SI{30}{\degree}$.  The results are plotted for four different wavelengths: \SI{1.50}{\micro\metre} (a), \SI{1.55}{\micro\metre} (b), \SI{1.60}{\micro\metre} (c), and \SI{1.65}{\micro\metre} (d).  A regular sampling has been performed on the parameter regions ranging from $w_e=$\SIrange{0.4}{2.6}{\micro\metre} and $R_e=$\SIrange{16}{19.5}{\micro\metre}.  The shaded contours deline the parameter regions where a positive improvement is obtained. At the low $R_e$ edge, there is a region showing a very steep enhancement of radiation as $R_e$ decreases.  Above it, the improvement factor shows a highly oscillatory behavior both when $R_e$ and $w_e$ are varied, with local maxima diminishing their amplitude as the value of $w_e$ grows and, even more strongly, as $R_e$ is increased. There are little variations with wavelength in the range of $\lambda_0$ from \SIrange{1.55}{1.65}{\micro\metre}, but the differences are noticeable for the results at the largest $R_e/\lambda_0$ shown at Fig. \ref{fig::exteriorR15} (a).  In this plot, the peak values are slightly reduced and there is also a reduction in the shaded surface corresponding to the positive loss improvement regions.  The domain around the global maxima for all the spectral range, near ($R_e=\SI{17}{\micro\metre}$,$w_e=\SI{0.5}{\micro\metre}$), is a reasonable working point also when fabrication tolerances are taken in consideration.

\begin{figure}[h]
\centering
\begin{tabular}{cc}
{\large (a)}&{\large (b)}\\
\includegraphics[width=2.5in]{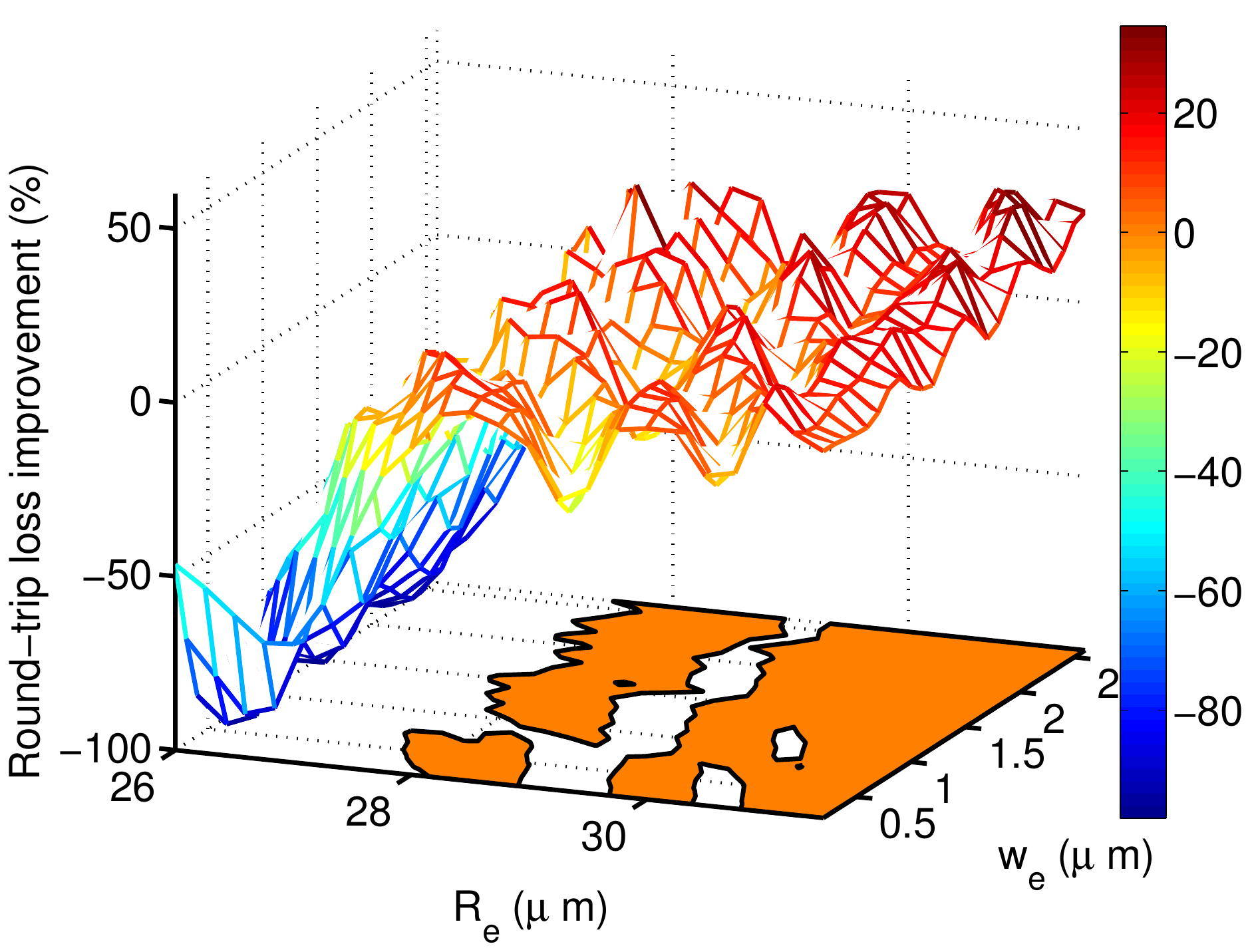}&\includegraphics[width=2.5in]{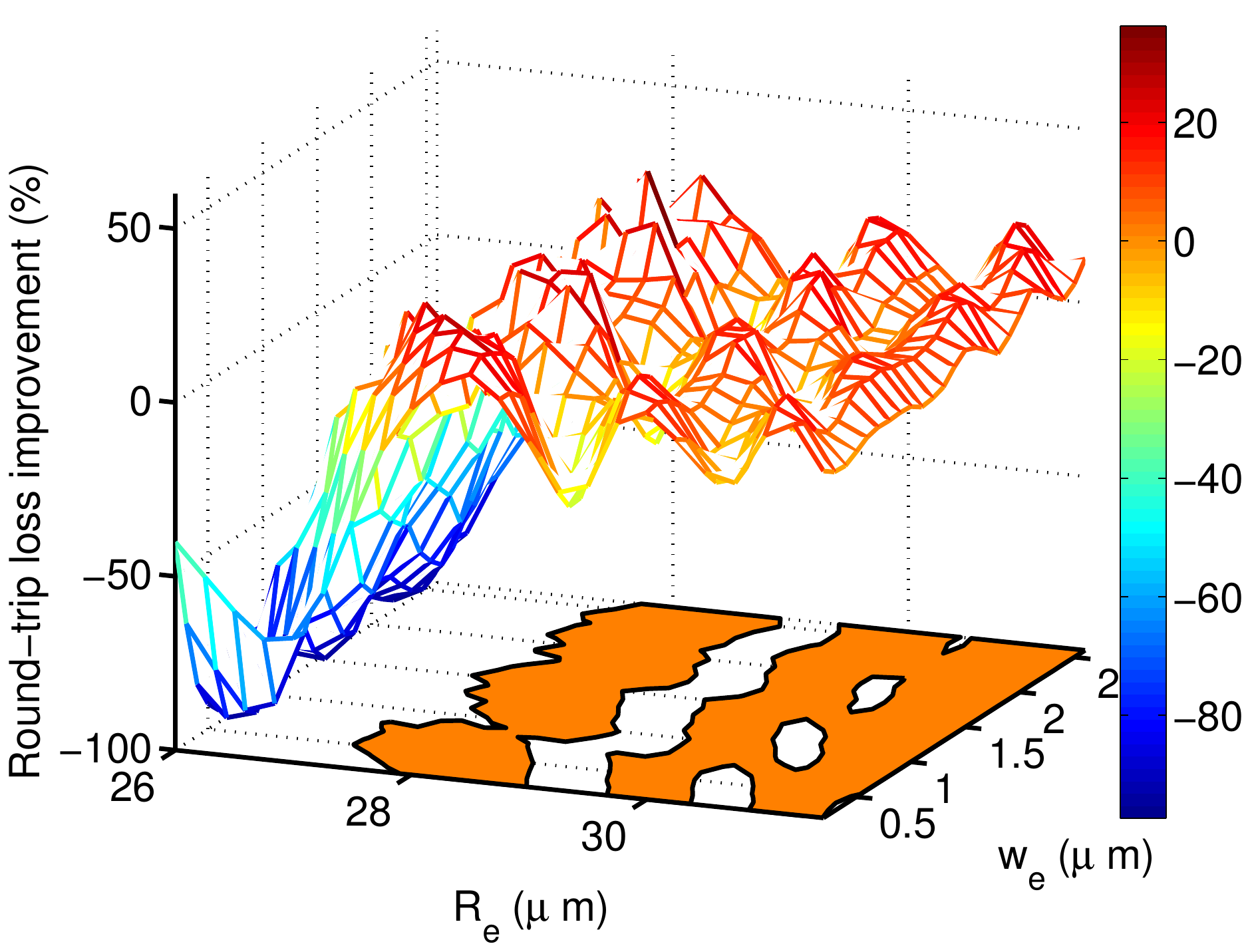}\\
{\large (c)}&{\large (d)}\\
\includegraphics[width=2.5in]{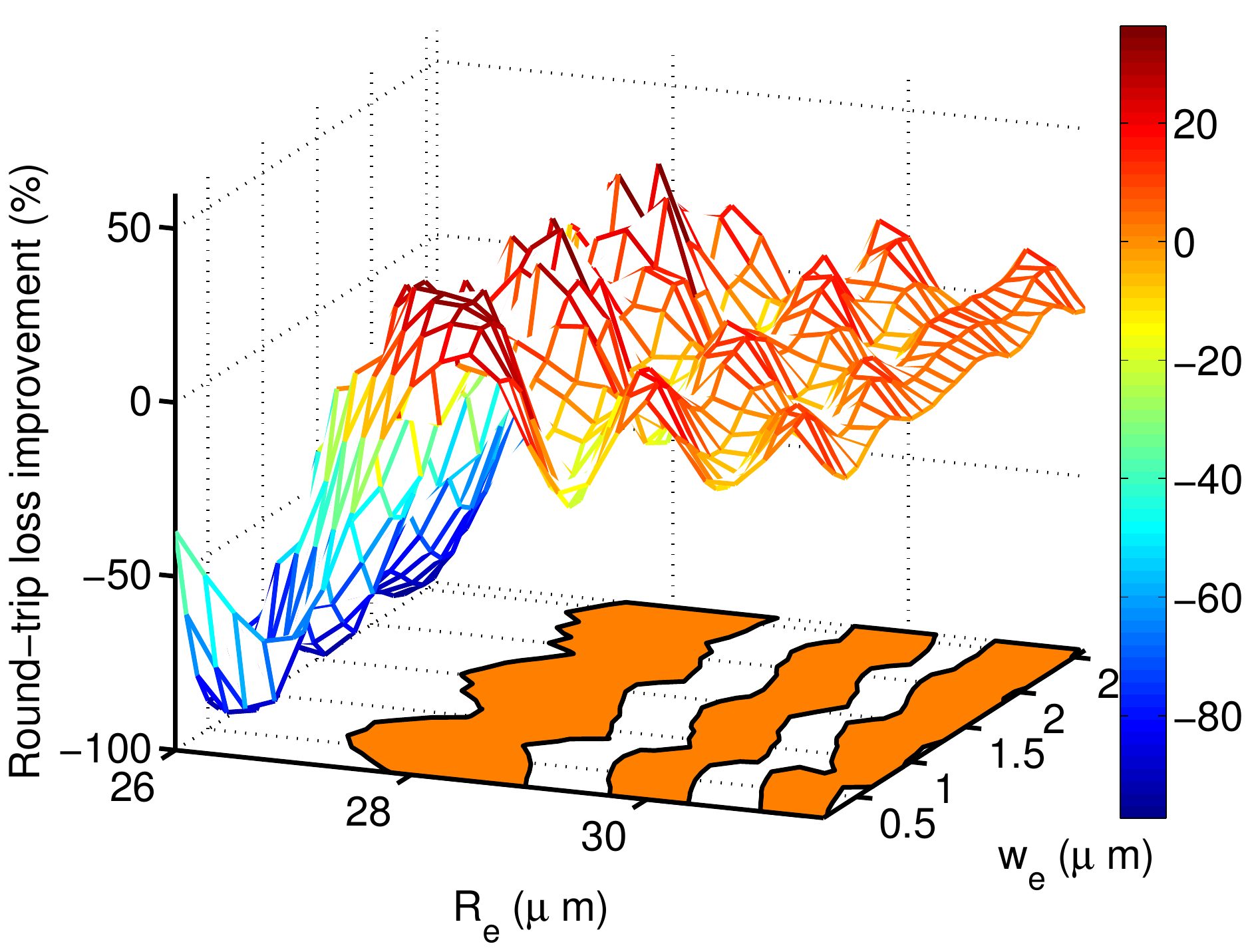}&\includegraphics[width=2.5in]{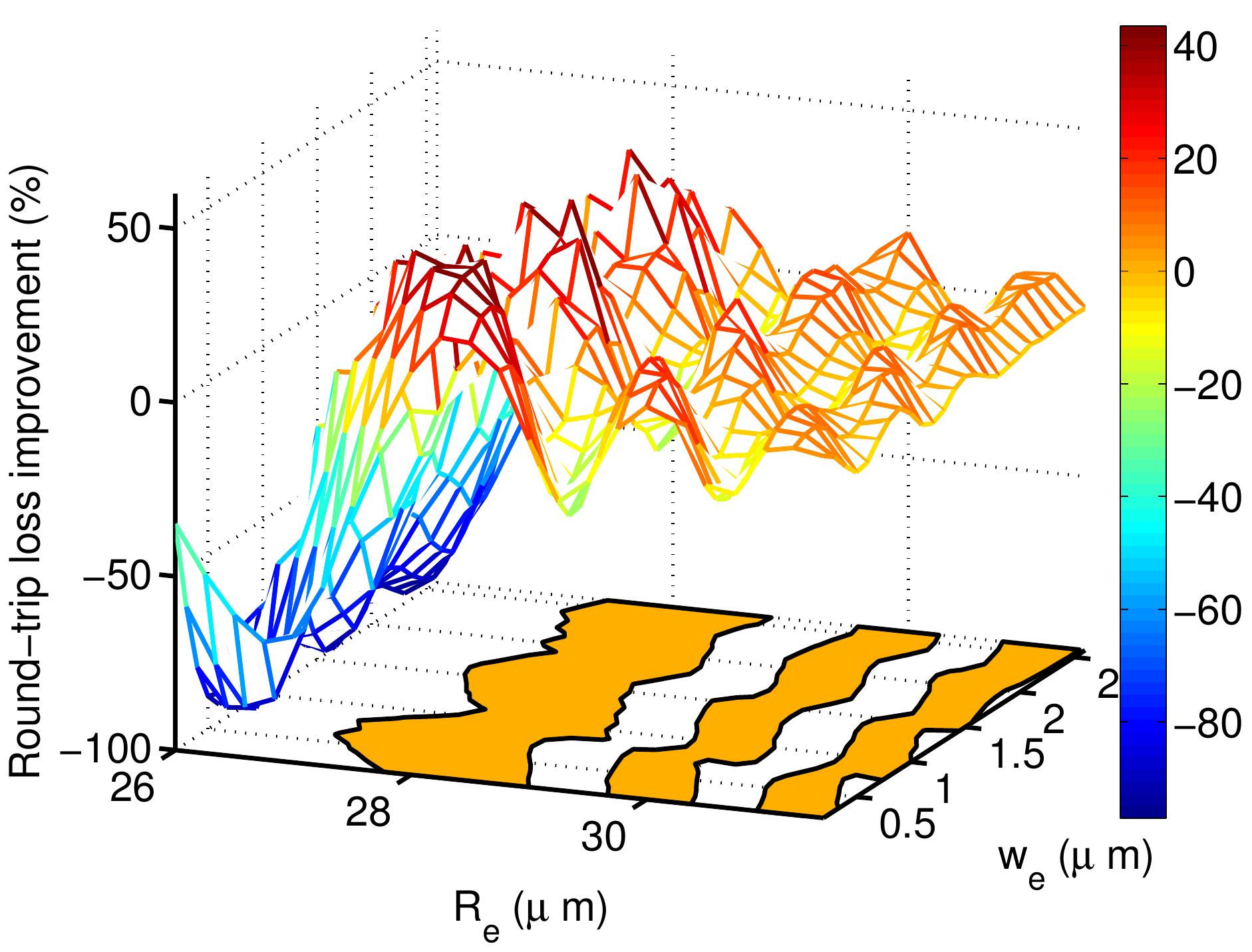}
\end{tabular}
\caption{Round-trip loss improvement factors as a function of the exterior annuli parameters $R_e$ and $w_e$ for fixed $R=$\SI{25}{\micro\metre}  $\theta_1=\theta_2=\theta=\SI{30}{\degree}$.  Four vacuum wavelength values are considered: (a) $\lambda_0$=\SI{1.50}{\micro\metre}, (b) $\lambda_0$=\SI{1.55}{\micro\metre}, (c) $\lambda_0$=\SI{1.60}{\micro\metre} and (d) $\lambda_0$=\SI{1.65}{\micro\metre}.}
\label{fig::exteriorR25}
\end{figure}

Fig. \ref{fig::exteriorR25} displays the improvement factors for a racetrack with $R=$\SI{25}{\micro\metre} and $\theta=\SI{30}{\degree}$.  Figs. \ref{fig::exteriorR25} (a), (b), (c) and (d) correspond, respectively, to $\lambda_0=$\SI{1.50}{\micro\metre} (a),  $\lambda_0=$\SI{1.55}{\micro\metre} (b),  $\lambda_0=$\SI{1.60}{\micro\metre} (c), and  $\lambda_0=$\SI{1.65}{\micro\metre} (d). In this case, the parameter region studied ranges from $w_e=$\SIrange{0.2}{2.6}{\micro\metre} and $R_e=$\SIrange{26}{31.5}{\micro\metre}.   The oscillatory behavior of the results presented in Fig. \ref{fig::exteriorR25} are similar to those in Fig. \ref{fig::exteriorR15}.  Nevertheless, the tendency in the decrease of peak values of the improvement  factor as $R/\lambda_0$ increases, already observed in Fig. \ref{fig::exteriorR15} (a), is now more evident.  This is accompanied, as going from \ref{fig::exteriorR25} (d) to \ref{fig::exteriorR25} (a) with a decrease of the amplitude of the oscillations and the appearance of peaks of similar improvement factor at different values of $w_e$ for the fixed value of $R_e\simeq\SI{28}{\micro\metre}$.  This opens the possibility of other possible operation regions that will be further explored.  At the largest $R/\lambda_0$, the results displayed in Fig. \ref{fig::exteriorR25} (a) shows a shift of the peak values of the local maxima to larger values of both $R_e$ and $w_e$. 

Therefore, the overall picture depicted by Figures \ref{fig::exteriorR15} and \ref{fig::exteriorR25} shows systematic variations with  $R/\lambda_0$ of the radiation-quenching properties provided by the external ring sector that are similar to those observed for the lateral offset implementation in a certain frequency range, whereas the improvement does not vary significantly as $R/\lambda_0$ is sufficiently small. In any case, the operation close to a selected local maximum is not compromised by fabrication tolerances due to smoothness of the variation of the improvement factor with the parameters in all cases.

The inspection of  Figs. \ref{fig::offset}, \ref{fig::exteriorR15} and \ref{fig::exteriorR25} shows that comparable reduction factors in the radiation loss can be  obtained by the two optimization strategies.  For the shorter radius, with larger radiation losses at the bent sections, the improvement provided by the radiation quenching annulus is larger than that of the lateral offset, and the converse situation is found for the larger radius racetrack.  For the simulations performed, it is observed an increase of the improvement obtained with the lateral offset and a simultaneous decrease  (for a certain frequency range) of that of the external ring sector as $R/\lambda_0$ grows.

We now address the effect of the variation of the improvement factor with $\theta$.  The study is performed at three different operation points.  The first, for $R=$\SI{15}{\micro\metre}, is near the largest local maximum in Fig. \ref{fig::exteriorR15}, $R_e=$\SI{17}{\micro\metre} and $w_e=$\SI{400}{\nano\metre}.  For $R=$\SI{25}{\micro\metre}, two sets of values are selected.  The first is at $R_e=$\SI{28}{\micro\metre} and $w_e=$\SI{2.2}{\micro\metre}, close to a maximum obtained for relatively large values of $R_e$.  The second corresponds to the vicinity of the peak with the smallest $R_e$ and $w_e$, similar to the maximum considered for $R=$\SI{15}{\micro\metre}, with $R_e=$\SI{28}{\micro\metre} and $w_e=$\SI{400}{\nano\metre}.

\begin{figure}[H]
\centering
\begin{tabular}{cc}
{\large (a)}\\
\includegraphics[width=3in]{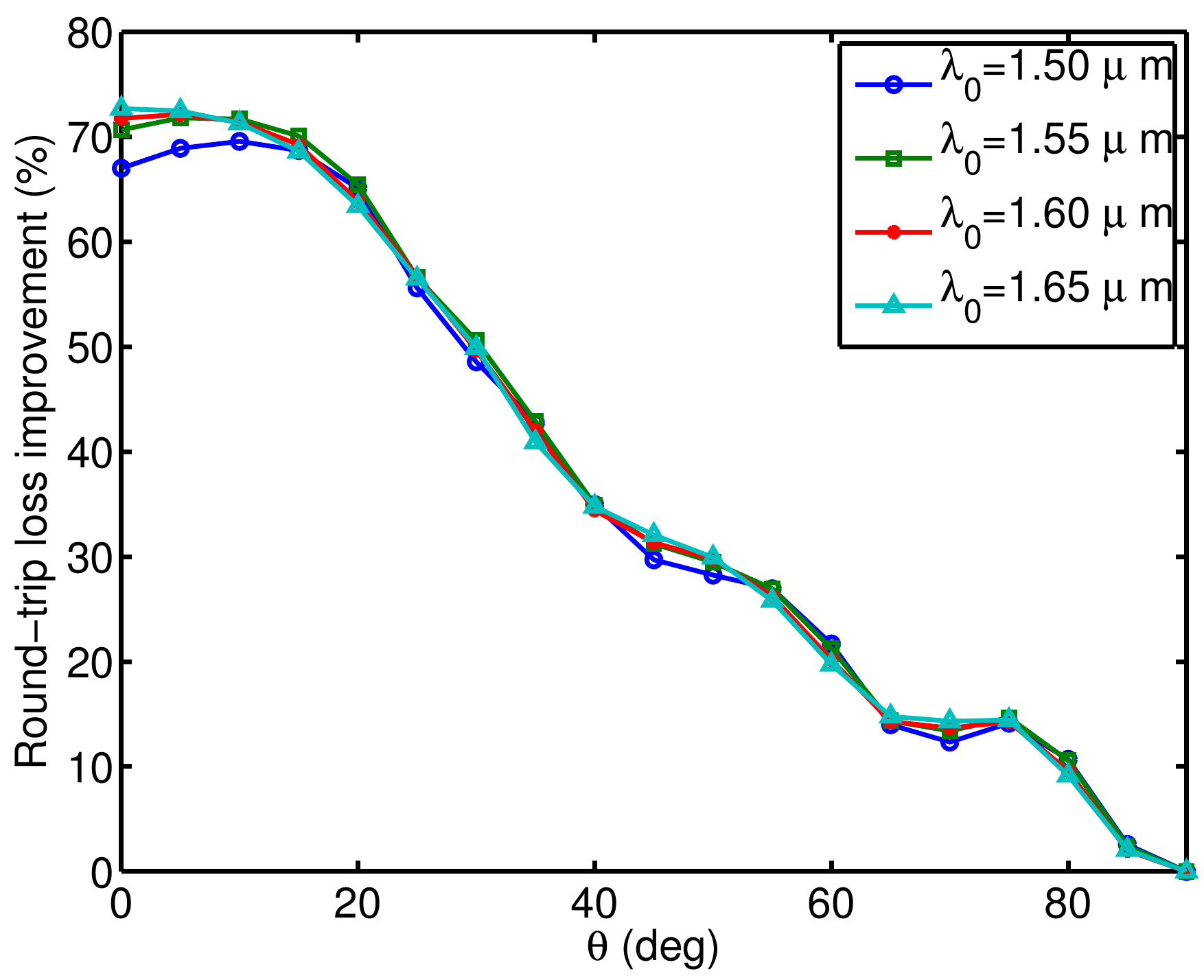}\\
{\large (b)}\\
\includegraphics[width=3in]{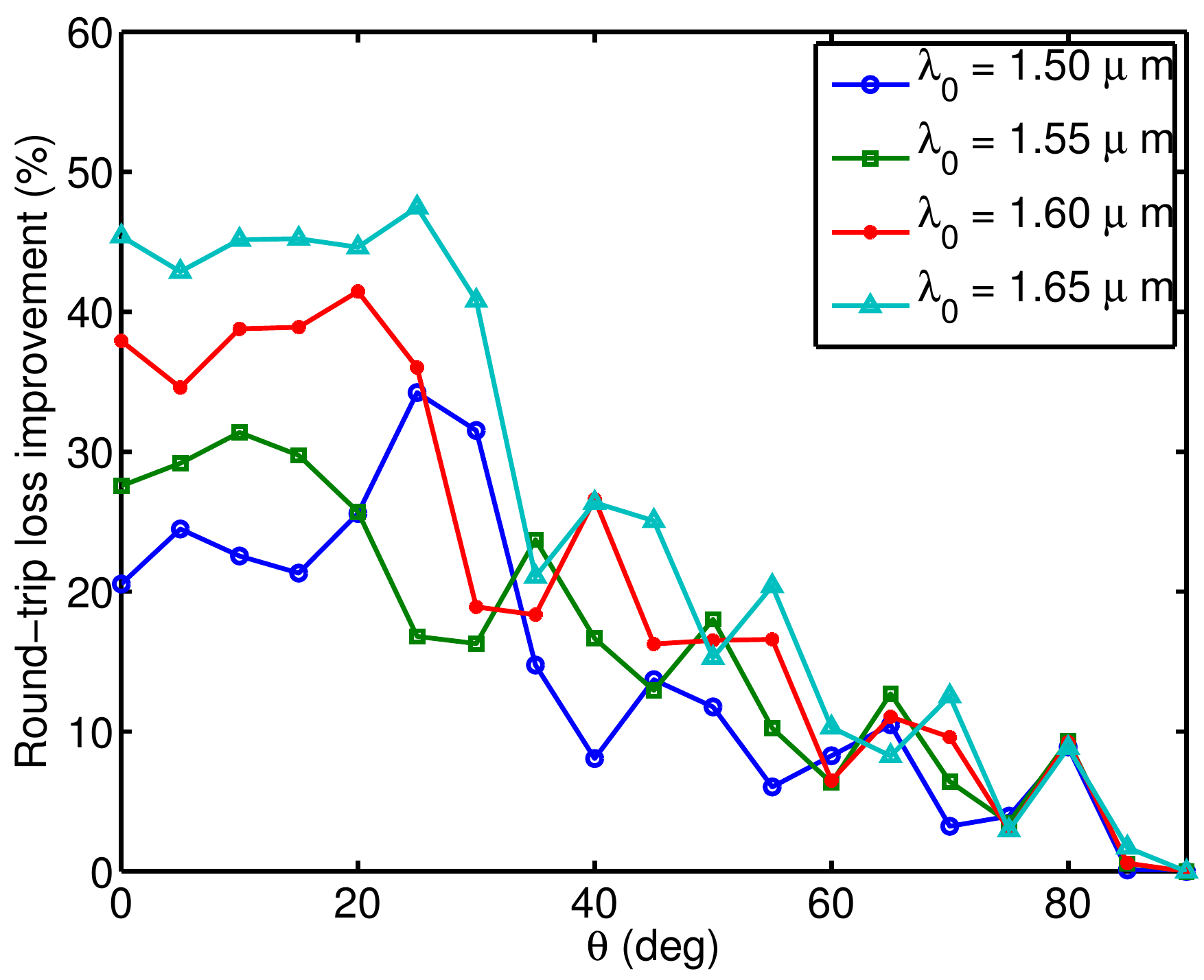}\\
{\large (c)}\\
\includegraphics[width=3in]{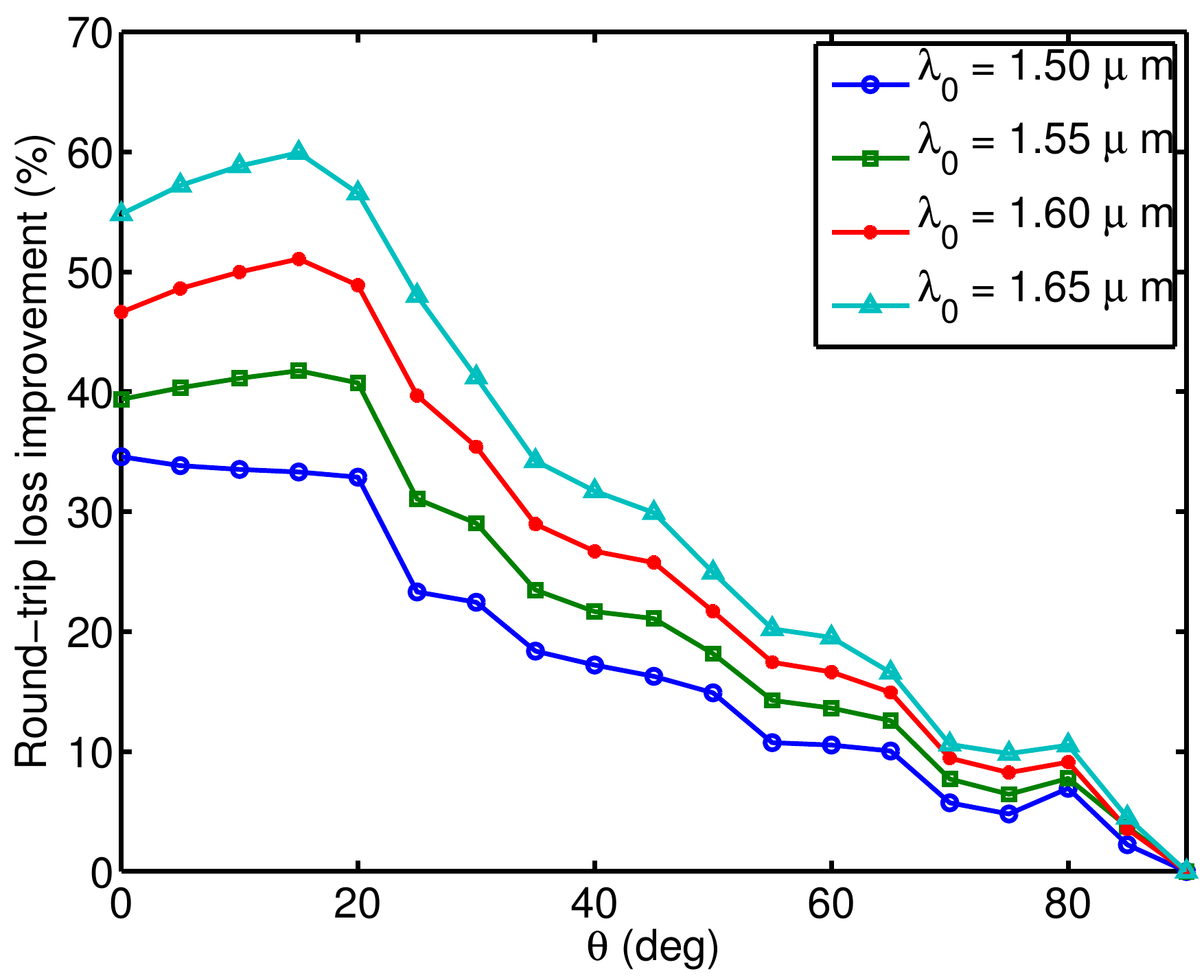}
\end{tabular}
\caption{Variation of the round-trip loss improvement factor with $\theta$ for three different sets of parameters: (a) $R=$\SI{15}{\micro\metre},  $R_e=$\SI{17}{\micro\metre} and $w_e=$\SI{400}{\nano\metre}, (b) $R=$\SI{25}{\micro\metre}, $R_e=$\SI{28}{\micro\metre} and $w_e=$\SI{2.2}{\micro\metre}, (c) $R=$\SI{25}{\micro\metre}, $R_e=$\SI{28}{\micro\metre} and $w_e=$\SI{400}{\nano\metre}}
\label{fig::theta}
\end{figure}

The improvement factors for the three aforementioned cases and four wavelengths covering the band between \SIrange{1.50}{1.65}{\micro\metre} are shown in Fig. \ref{fig::theta}.  It would be expected an enhancement as $\theta$ decreases, providing a broader coverage of the bent waveguide.  This is, in general, the case but some remarks are due. As the coverage is close to complete when $\theta$ approaches zero, there is a slow-down of the increase of the improvement factor with decreasing $\theta$ or even a decrease.  This can be attributed to the interaction with the discontinuity between the straight and bent waveguides.  Also, the results in Fig. \ref{fig::theta} (b) are clearly distinct from those of Figs. \ref{fig::theta} (a) and (c), showing wide variations of the improvement factor with the value of $\theta$, although the global tendency of a reduction with an increase in $\theta$ is in general preserved.  It is also important to note the differences observed in the dependence with the optical wavelength.  Whereas for $R=$\SI{15}{\micro\metre} the variations with frequency in the whole band are minimal, very wide wavelength-dependent variations can be observed in the results for $R=$\SI{25}{\micro\metre}.   This is consistent with the frequency dependent behavior observed in Fig. \ref{fig::exteriorR15} and Fig. \ref{fig::exteriorR25}. It is also observed that the changes with frequency are more important as the value of $\theta$ decreases. Even though the improvement factors obtained at the two maxima in Fig. \ref{fig::exteriorR25} that correspond to the results in Figs.  \ref{fig::theta} (b) and (c), the wider variations with $\theta$ in the case of Fig.  \ref{fig::theta} (b) can be associated with a better stability of the local maximum at the smallest values of $R_e$ and $w_e$.

\subsection{$Q$ factor}

We now address, by numerical simulation, the quantitative effect of the proposed geometry modifications both in the intrinsic and loaded $Q$ factor of the racetrack resonators.  The length of the straight resonator sections is $L_s=$\SI{30}{\micro\metre} in all cases, and two different values of $R$ are considered, $R=$\SI{15}{\micro\metre} and $R=$\SI{25}{\micro\metre}.

\begin{figure}[H]
\centering
\begin{tabular}{c}
\begin{tabular}{cc}
{\large (a)}&{\large (b)}\\
\includegraphics[width=1in]{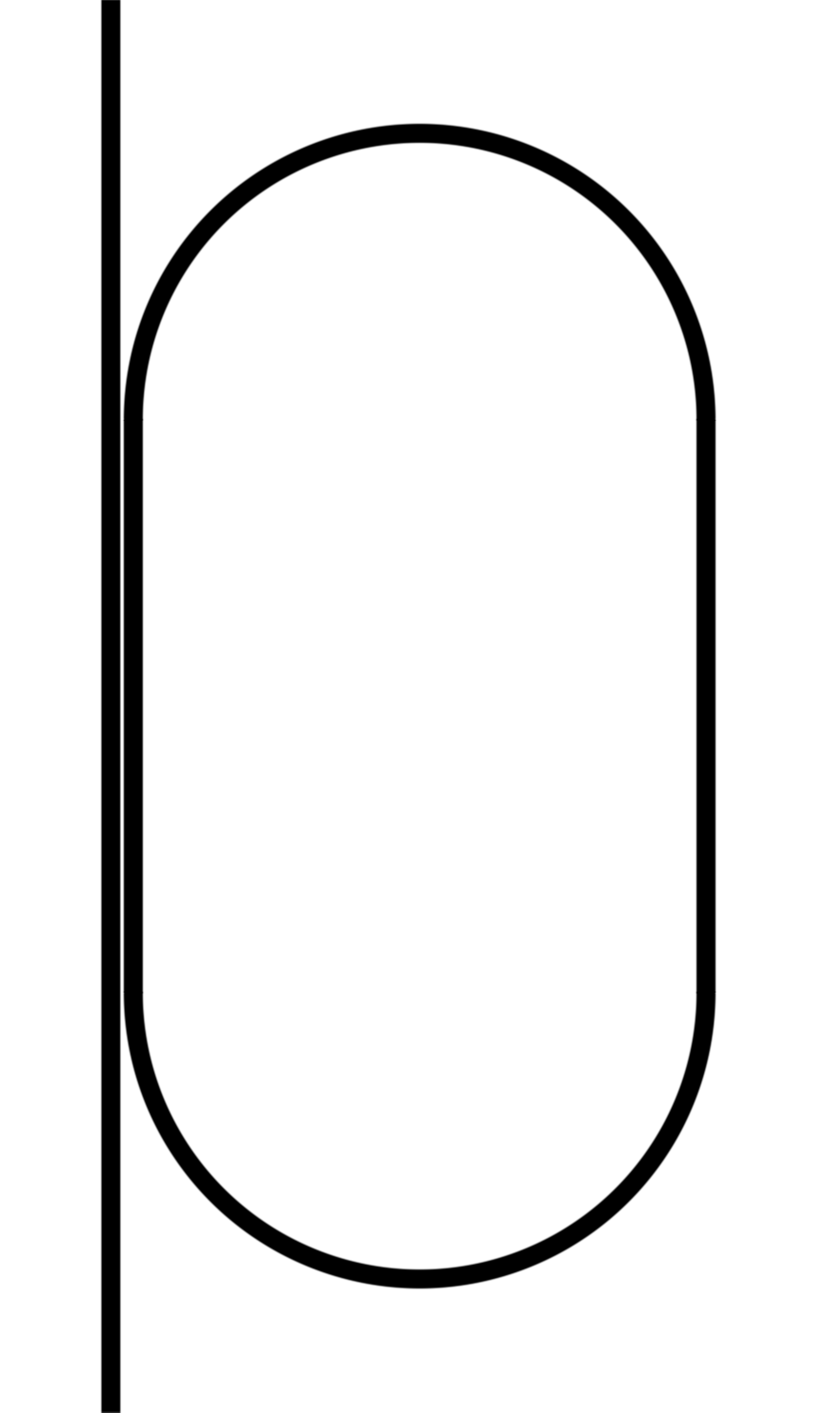}&
\includegraphics[width=1in]{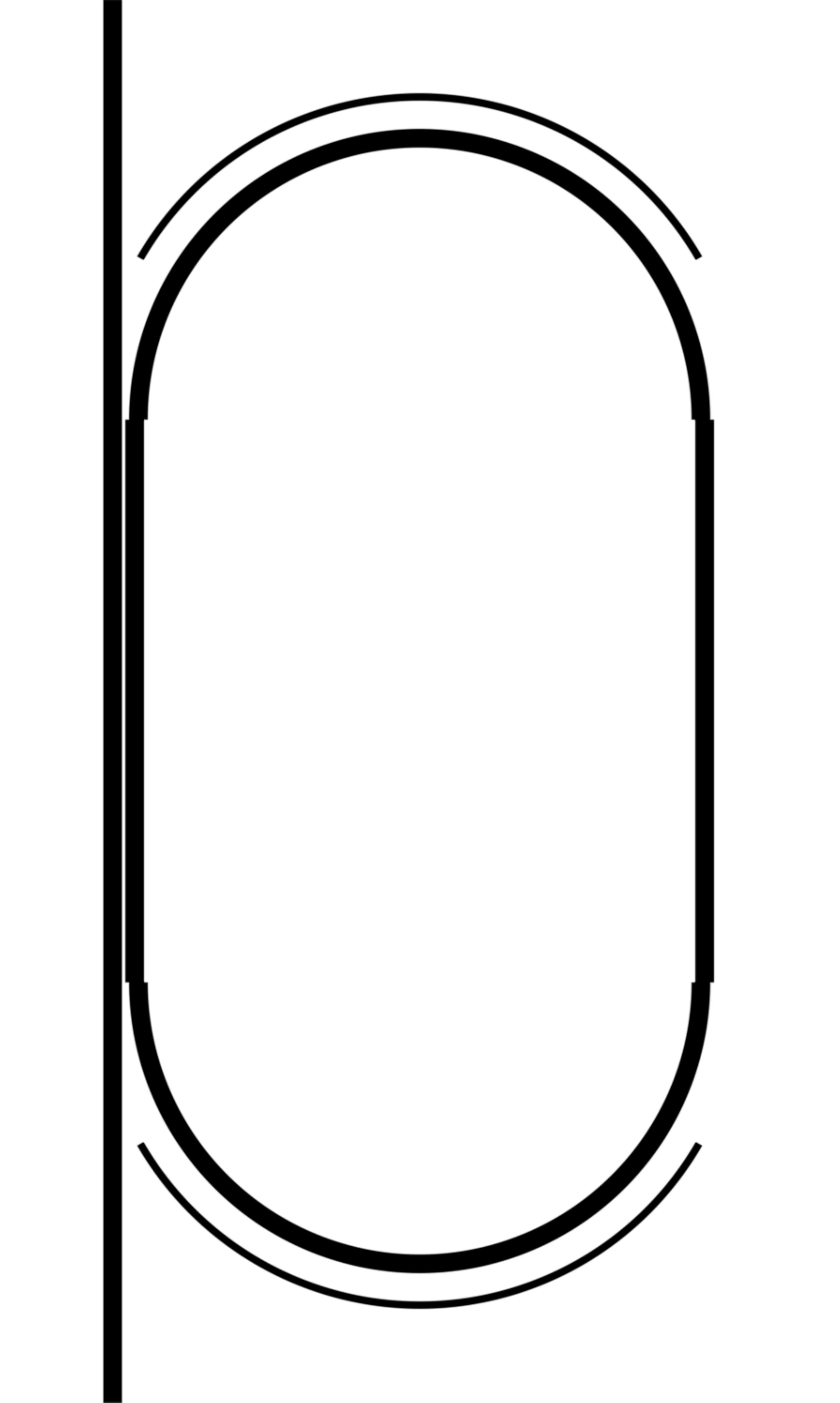}\\
\end{tabular}\\
\begin{tabular}{c}
{\large (c)}\\
\includegraphics[width=3in]{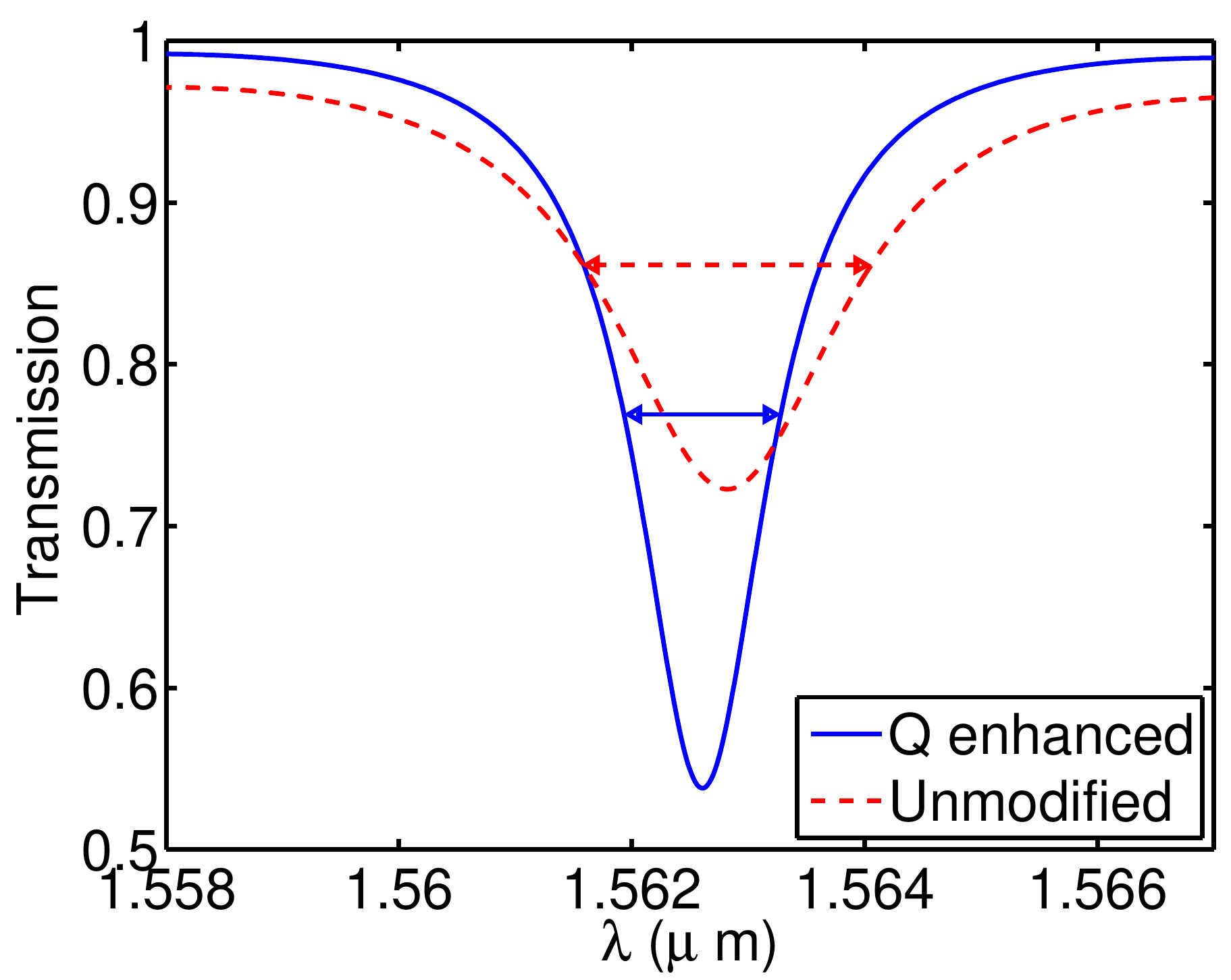}\\
{\large (d)}\\
\includegraphics[width=3in]{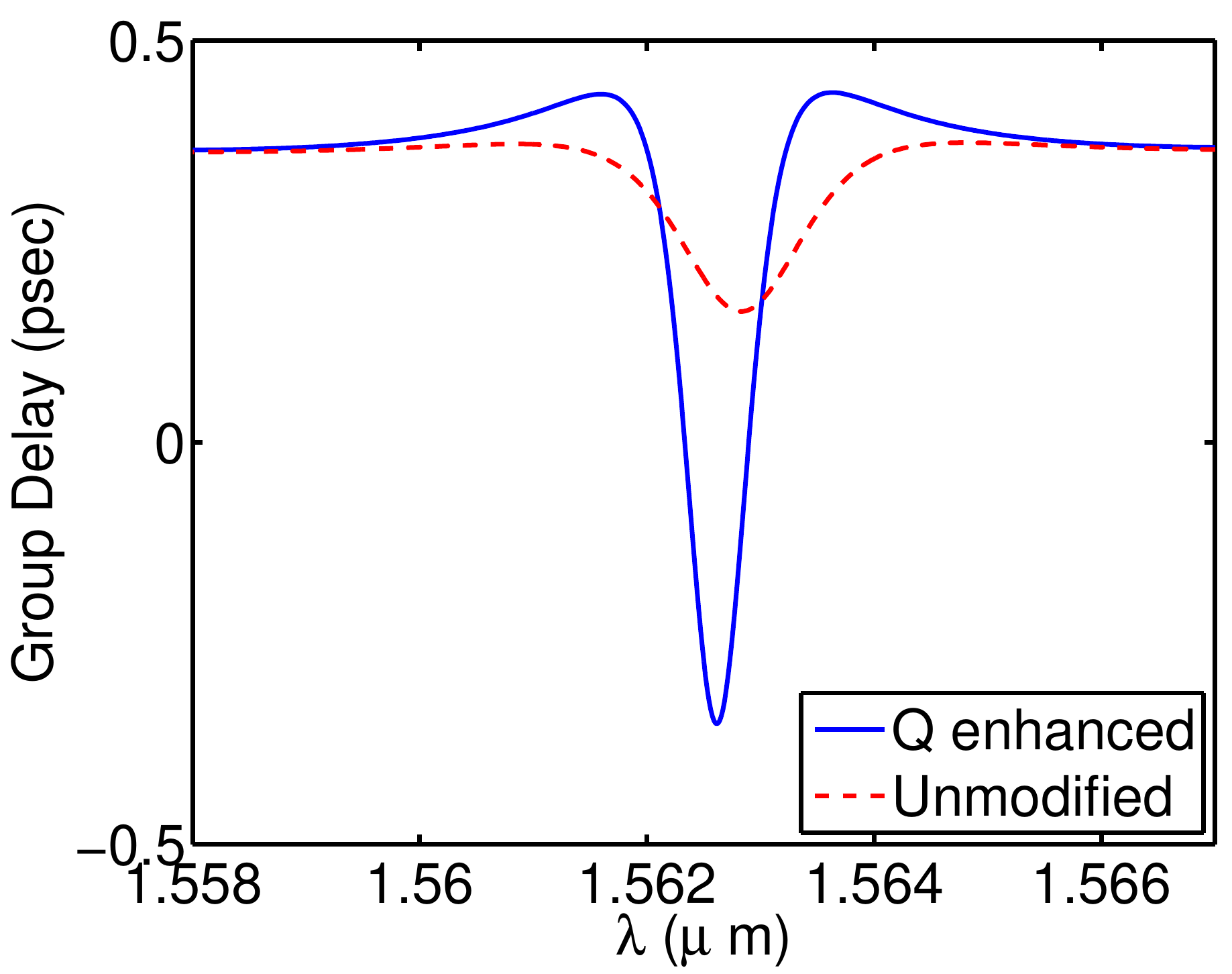}
\end{tabular}
\end{tabular}
\caption{Transmission (c) and group delay (d) for a racetrack resonator with $R=$\SI{15}{\micro\metre} side coupled to a straight waveguide section at the resonance near $\lambda=$\SI{1.56}{\micro\metre}.  The results for the conventional geometry (a) are shown with dashed lines and those of the $Q$-enhanced version (b) with solid lines. Arrows label the respective FWHM bandwidths.}
\label{fig::transmision}
\end{figure}

We will compare the resonator $Q$ with and without radiation loss mitigation measures for both values of $R$. In all cases, we consider $\theta=\theta_1=\theta_2=$\SI{30}{\degree}, that is compatible with the introduction of access waveguides, as shown in Fig. \ref{fig::transmision} (b). The previously determined optimal values of $l_{off}$ are used, that correspond to $l_{off}=$\SI{200}{\nano\metre} and  $l_{off}=$\SI{90}{\nano\metre}  for $R=$\SI{15}{\micro\metre} and $R=$\SI{25}{\micro\metre}, respectively.

As regards the external annuli, the same three set of values corresponding to local maxima in the $(R_e,w_e)$ space that were analyzed regarding the dependence with $\theta$ will be used.  These correspond to $(R_e=\SI{17}{\micro\metre},w_e=\SI{400}{\nano\metre})$ for  $R=$\SI{15}{\micro\metre}, and to  $(R_e=\SI{28}{\micro\metre},w_e=\SI{2.2}{\micro\metre})$ and $(R_e=\SI{28}{\micro\metre},w_e=\SI{400}{\nano\metre})$ for  $R=$\SI{25}{\micro\metre}.

\begin{figure}[H]
\centering
\begin{tabular}{cc}
{\large (a)}\\
\includegraphics[width=3in]{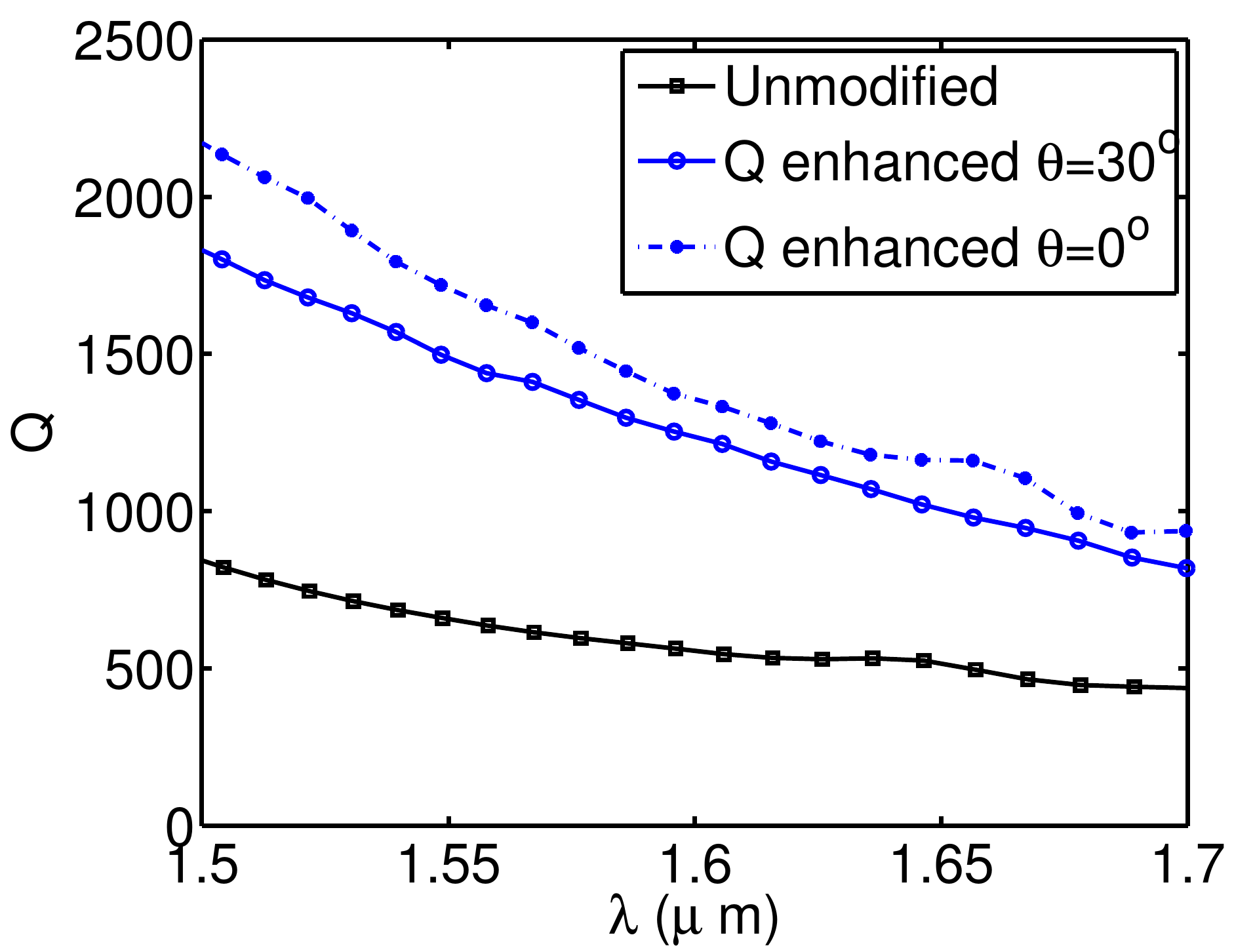}\\
{\large (b)}\\
\includegraphics[width=3in]{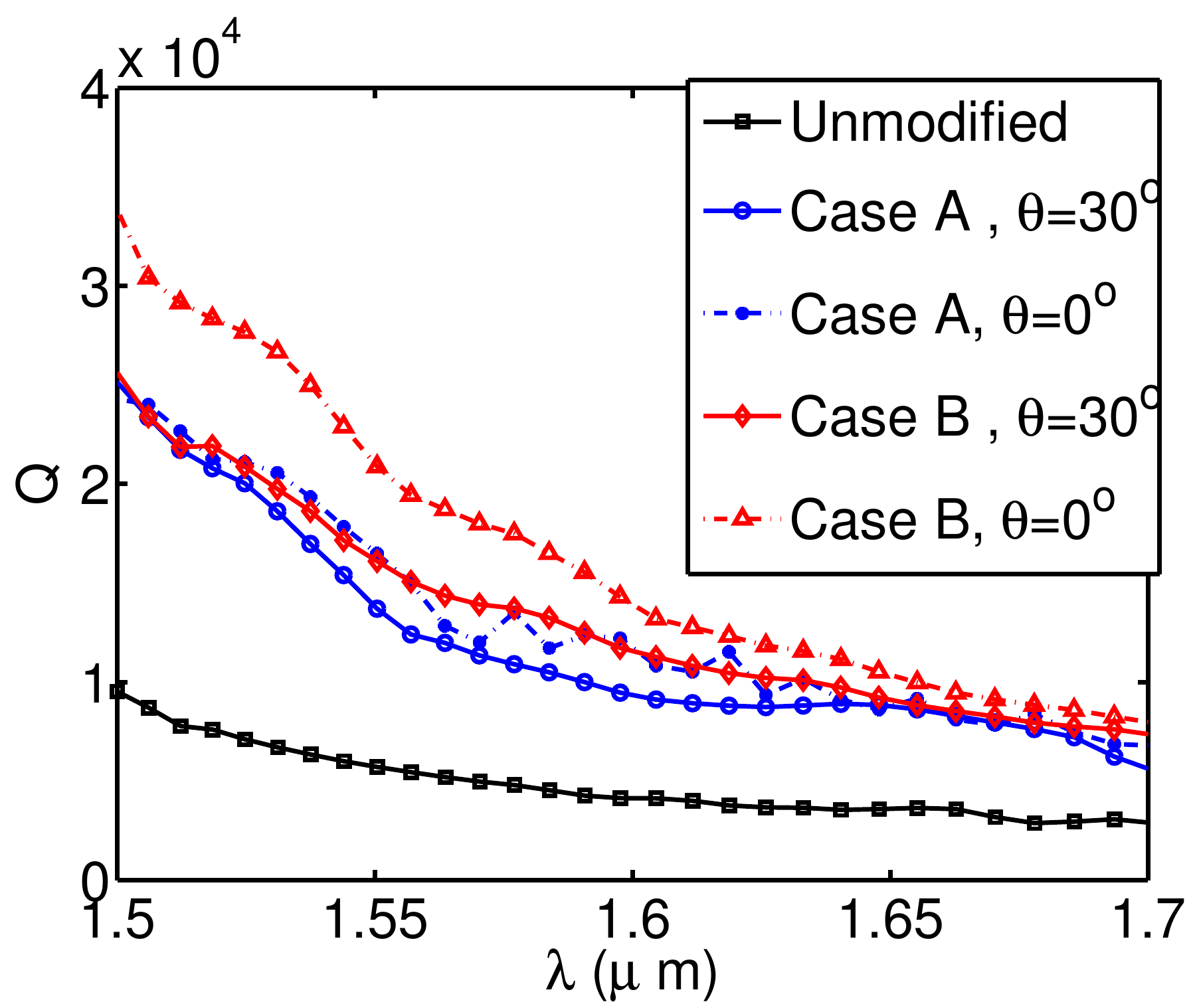}
\end{tabular}
\caption{Unloaded $Q$ factors calculated for (a) $R=\SI{15}{\micro\metre}$ and (b) $R=\SI{25}{\micro\metre}$ microresonators.  Squares correspond to conventional geometries.}
\label{fig::Q}
\end{figure}

\subsubsection{Unloaded $Q$}

The racetrack microresonators have been characterized in an unloaded standalone configuration, without any coupling to an external circuit, by injecting an input signal in the upper straight section of the resonator and determining the frequencies and decay rates of the modes from the temporal evolution of the field at the opposite side of the resonator.  The filter diagonalization method \cite{harminv} has been employed to determine the resonances and their respective $Q$-factors.  

The calculated results in the spectral region between \SI{1.5}{\micro\metre} and \SI{1.7}{\micro\metre} are shown in Figures \ref{fig::Q} (a) and (b) for $R=$\SI{15}{\micro\metre} and $R=$\SI{25}{\micro\metre}, respectively.  In panel (b), case A corresponds to $(R_e=\SI{28}{\micro\metre},w_e=\SI{2.2}{\micro\metre})$ and case B to $(R_e=\SI{28}{\micro\metre},w_e=\SI{400}{\nano\metre})$. Resonance frequencies and $Q$-factors for the conventional racetrack geometry are shown with squares.  The results for the modified geometry proposed in this work show an improvement of the $Q$-factor by over $100\%$ in most cases.  As expected from the optimization previously performed, cases A and B display comparable improvement factors, even though case B is better if the whole optimization band is considered.  For the three cases, the geometry with $\theta_1=\theta_2=\SI{0}{\degree}$ has also been studied.  The results show the expected improvement in the $Q$-factor, consistent with the calculations of the previous subsection. 

The values of the $Q$-factor illustrated in the calculations could be increased further, at the cost of reducing the free spectral range, simply by enlarging the straight waveguide sections since the $Q$-factor scales linearly with the total ring length \cite{bogaerts} when losses are kept constant.  This would be unpractical in the presence of high intrinsic material losses, such as in the SOI platform, but not in other technologies with lower intrinsic losses, like silicon nitride.

\subsubsection{Loaded $Q$} 

The loaded $Q$-factors of the structures under analysis have been also addressed by studying the transmission properties of a $R=\SI{15}{\micro\metre}$ racetrack microresonator side coupled to a bus waveguide with and without the $Q$-enhancing geometry modifications.  The loaded $Q$-factor for a side coupled racetrack resonator at a resonance wavelength $\lambda_{res}$  is given by \cite{bogaerts}
\begin{equation}
Q_L=\dfrac{\pi n_g L \sqrt{ra}}{\left(1-ra\right)\lambda_{res}}\label{eq::QL}
\end{equation}
where $L$ is the ring length, $a$ is the round-trip field amplitude loss factor and $r$ is the mark field amplitude transmission coefficient of the coupler. The corresponding FWHM bandwidth is
\begin{equation}
\Delta\lambda_\text{FWHM}=\dfrac{\lambda_{res}}{Q_L}.\label{eq::QL2}
\end{equation}
Therefore, at a given wavelength, an increase of the value of $Q_L$ corresponds to a reduction of the FWHM bandwidth by the same factor.

According to \eqref{eq::QL},  the same value of $r$ has to be used for a fair comparison between the structures.  Nevertheless, the value of the effective coupler length is affected by the interaction between the field in the access and resonator waveguides in the curved sections close to the coupler  \cite{xia} in a different manner when the lateral offset is introduced.  Therefore, in order to obtain approximately the same coupling coefficient $r$ in both cases, the separations between the two coupler waveguides for the unmodified and modified geometries have been set to \SI{220}{\nano\metre} and \SI{190}{\nano\metre}, respectively.  The same width of \SI{1}{\micro\metre} used in the ring has been assumed for the bus waveguide.  The two geometries are shown in Figures \ref{fig::transmision} (a) and (b).  The transmission function at a resonance near $\lambda=$\SI{1.56}{\micro\metre} is shown in Figure \ref{fig::transmision} (c) and the corresponding transmission group delay in Figure \ref{fig::transmision} (d).  The results for our design are shown with solid lines and those of the conventional geometry with dashed lines. Changing the coupler waveguide separation produces a small relative shift of the resonances.   The data obtained shows, on the other hand, a deeper and narrower resonance for the $Q$- enhanced geometry, consistently with the significantly larger values of the intrinsic unloaded $Q$ previously determined.  For the unmodified geometry, the resonance FWHM bandwidth is $\Delta\lambda_\text{FWHM}=$\SI{2.5}{\nano\metre} and it is reduced with the proposed geometry to $\Delta\lambda_\text{FWHM}=$\SI{1.3}{\nano\metre}.

\section{Conclusion}

In this work, a modified racetrack microresonator design has been proposed and analyzed.  The new geometry aims to reduce the radiation losses, while fully keeping the original versatile connectivity properties of the device as part of integrated optical circuits.  The proposal is particularly interesting for integrated photonics platforms with radiation losses due to curvatures potentially dominating the total propagation loss, such as is the case of silicon nitride.

Computer simulations of loaded and unloaded resonators show the expected increase of the $Q$-factor in the new optimized geometries.  Even though device sizes have been kept relatively small in order to bound the computation times, the quality factor scales linearly with the total ring length \cite{bogaerts}, and larger $Q$-factors should be easily attainable at the cost of reducing the free spectral range.  This is an advantage in the implementation of racetrack microresonators with ultra-low loss optical integration platforms, such as Si\textsubscript{3}N\textsubscript{4}/SiO\textsubscript{2}, where the attenuation in the straight waveguide segments is very small and the radiation due to bent sections can be largely mitigated with the scheme proposed in this work. 

 In the optimization a value of $\theta_1=\theta_2=$\SI{30}{\degree} has been used in order to allow for the coupling to waveguides from both sides of the racetrack microresonator.  An intermediate design alternative to the symmetric sectoring of the radiation quenching slabs could be to use $\theta_1\ne$\SI{0}{\degree} and $\theta_2=\SI{0}{\degree}$.  This could be used in side coupled racetrack resonators.  For a racetrack in an add-drop configuration or as part of coupled resonator optical waveguides, the symmetric sectoring seems the most reasonable option.

The calculations performed show a comparable contribution to the enhancement of the $Q$ factor from the two geometry modifications.  Nevertheless, the dominant contribution comes from the lateral offset at the larger radius, whereas the converse is true at the shorter radius.

The analysis of the structures provides a guide for the determination of the optimal value of the $l_{off}$ parameter, which is only dependent on the radius of the bent waveguide sections and is not different from former calculations regarding the junction between straight and bent waveguide sections \cite{kitoh}.  As regards the design of the pulley ring sector, working at the local maximum obtained for the smallest values of $R_e$ and $w_e$ seems to be, in general, the option providing best performance and most stable operation for most the of the spectral range at bend radii considered.  Nevertheless, the results obtained show that this changes as $R/\lambda_0$ becomes larger.  Finally, a strategy based on the maximization of the value of $\theta$ within the constraints imposed by the requirement of coupling to the external circuit can be advised from the results of the calculations performed in this work. 

It is noteworthy that the same design considerations used for the control of the radiation loss in the curved sections can be applied also to microring resonators, limiting the radiation reduction slab to an angular sector such that the normal evanescent coupling of the ring resonator to a side structure is conveniently permitted. 

\section*{Acknowledgments}

This work has been was supported by Junta de Castilla y Le\'on, Project No. VA089U16, and the Spanish Ministerio de Econom\'{\i}ıa y Competitividad, Project No. TEC2015-69665-R.


\begin{thebibliography}{99}

\bibitem{little1} B.E. Little, S.T. Chu, H.A. Haus, J. Foresi, and J.-P. Laine, ``Microring Resonator Channel Dropping Filters,'' J. Lightwave Technol. 15 (1997) 998-1005.

\bibitem{little2} B.E. Little, S.T. Chu, P.P. Absil, J.V. Hryniewicz, F.G. Johnson, F. Sieferth, D. Gill, V.Van, O. King and M. Trakalo, ``Very High-Order Microring Resonator Filters for WDM Applications,'' IEEE Photon. Technol. Lett. 16 (2004) 2263--2265. 

\bibitem{almeida} V.R. Almeida and M. Lipson, ``Optical bistability on a silicon chip,'' Opt. Lett. 29 (2004) 2387--2389.

\bibitem{vos} K. De Vos, J. Girones, S. Popelka, E. Schacht, R. Baets, and P. Bienstman, ``SOI optical microring resonator with poly(ethylene glycol) polymer brush for label-free biosensors applications,'' Biosens. Bioelectron. 24 (2009) 2528--2533.

\bibitem{xu} Q. Xu, B. Schmiddt, s. Pradhan, and M. Lipson, ``Micrometre-scale silicon electro-optic modulator,'' Nature 435 (2005) 325--327.

\bibitem{scheuer} J. Scheuer, G.T. Paloczi, J.K.S. Poon, and A. Yariv, ``Coupled resonator optical waveguides: Towards Slowing and Storing of Light,'' Opt. Photon. News 16 (2005) 36--40.

\bibitem{bogaerts} W. Bogaerts, P. DeHeyn, T. Van Vaerenbergh, K. De Vos, S. K. Selvaraja, T. Claes, P. Dumon, P. Bienstaman, D. Van Thourhout, and R. Baets, ``Silicon microring resonators,'' Laser Photonics Rev. 6 (2012) 47--73.

\bibitem{ou} H. Ou, K. Rottwitt, and H. Philipp, ``Deep glass etched microring resonators based on silica-on-silicon technology,'' Electron. Lett. 42 (2006) 581--583.

\bibitem{rabiei} P. Rabiei, W. H. Steier, C. Zhang, and L.R. Dalton, ``Polymero Micro-Ring Filters and Modulators,'' J. Lightwave Technol. 20 (2002) 1968--1975.

\bibitem{tien} M.C. Tien, J. F. Bauters, M.J.R. Heck, D. J. Blumenthal, adn J.E. Bowers, ``Ultra-low loss Si\textsubscript{3}N\textsubscript{4} waveguides with low nonlinearity and high power handling capability,'' Opt. Express 18 (2010) 23562--23568.

\bibitem{van} V. Van, T.A. Ibrahim, K. Ritter, P.P. Absil, F.G. Johnson,R. Grover, J. Goldhar,  P.-T. Ho, ``All-optical nonlinear switching in GaAs-AlGaAs microring resonators,'' IEEE Photonics Technol. Lett. 14 (2002) 74--76.

\bibitem{grover} R. Grover, P.P. Absil, V. Van, J.V. Hryniewicz, B.E. Little, O. King, L.C. Calhoun, F.G. Johnson, and P.-T. Ho, ``Vertically coupled GaInAsP-InP microring resonators,'' Opt. Lett. 26 (2001) 506-508. 



%
%
%


\bibitem{khurgin} J.B. Khurgin, ``Expanding the bandwidth of slow-light photonic devices based on coupled resonators,'' Opt. Lett. 30 (2005) 513--515. 2005.

\bibitem{haus} H.A. Haus, M.A. Popovi\'c, M.R. Watts, C. Manolatou, B.E. Little, and S.T. Chu, ``Optical resonators and filters,'' in \emph{Optical Microcavities}, K. Vahala, Ed.  World Scientific, 2004. pp. 1--37.


%
%

\bibitem{hosseini} E.S. Hosseini, S. Yegnanarayanan, A. H. Atabaki, M. Soltani, and A. Adibi, ``Systematic design and fabrication of high-Q single-mode pulley-coupled planar silicon nitride microdisk resonators at visible wavelengths,'' Opt. Express 18 (2010) 2127--2136.
 
\bibitem{hu} J. Hu, N. Carlie, N.N. Feng, L. Petit, A. Agarwal, K. Richardson, and L. Kimerling, ``Planar waveguide-coupled, high-index-contrast, high-Q resonators in chalcogenide glass for sensing,'' Opt. Lett. 33 (2008) 2500--2502.

\bibitem{cai} D.P. Cai, Y.H. Lu, C.C. Chen, C.C. Lee, C.E. Lin, T.J. Yen, ``High $Q$-factor microring resonator wrapped by the curved waveguide,'' Sci. Rep., vol. 5 (2015) 10078. 

\bibitem{kitoh} T. Kitoh, N. Takato, M. Yasu, and M. Kawachi, ``Bending loss reduction in Silica-Based Waveguides by Using Lateral Offsets,'' J. Lightwave Technol.  13 (1995) 555-562.


\bibitem{van2} V.Van, P.P. Absil, J.V. Hryniewicz, and P.-T. Ho, ``Propagation Loss in Single-Mode GaAs-AlGaAs Microring Resonators: Measurement and Model,'' J. Lightwave Technol. 19 (2001) 1734--1739. 



\bibitem{bauters} J.F. Bauters, M.J.R. Heck, D. John, D. Dai, M.-C. Tien, J.S. Barton, A. Leinse, R.G. Heideman, D.J. Blumenthal, and J.E Bowers, ``Ultra-low-loss high-aspect-ratio Si\textsubscript{3}N\textsubscript{4}/SiO\textsubscript{2} waveguides,'' Opt. Express 19 (2011) 3163--3174.


\bibitem{moss} D.J. Moss, R. Morandotti, A. L. Gaeta and M. Lipson, ``New CMOS-compatible platforms based on silicon nitride and Hydex for nonlinear optics,'' Nature Photonics 7 (2013) 597--607. 


\bibitem{xiong} C. Xiong, X. Zhang, A. Mahendra, J. He,D.-Y. Choi, C. J. Chae, D. Marpaung, A. Leinse, R. G. Heideman, M. Hoekman, G. H. Roeloffzen, R. M. Oldenbeuving, P. W. L. van Dijk, C. Taddei, P. H. W. Leong, and B. J. Eggleton, ``Compact and reconfigurable silicon nitride time-bin entanglement circuit,'' Optica 2 (2015) 724--727. 


%
%


\bibitem{yurtsever} G. Yurtsever, B. Povazay, A. Alex, B. Zbihian, W. Drexler, and R. Baets, ``Photonic integrated Mach-Zehner interferometer with an on-chip reference for optical coherence tomography,'' Biomed. Opt. Express 5 (2014) 1050--1061.

\bibitem{cai2} H. Cai, and A.W. Poon, ``Optical trapping of microparticles using silicon nitride waveguide junctions and tapered-waeguide junctions on an optofluidic chip,'' Lab Chip 12 (2010) 3803--3809.

\bibitem{ymeti} A. Ymeti, J.S. Kanger, J. Greve, G.A.J. Besselink, P.V. Lambeck, R. Wijn, and R.G. Heideman, ``Integration of microfluidics with a four-channel integrated optical Young interferomenter immunosensors,'' Biosens. Biolectron. 20 (2005) 1417--1421.


%



\bibitem{luke} K. Luke, A. Dutt, C.B. Potras, and M. Lipson, ``Overcoming Si\textsubscript{3}N\textsubscript{4}/SiO\textsubscript{2} film stress limitations for high quality factor ring resonators,'' Opt. Express 21 (2013) 22829--22833.

\bibitem{tien2} M-Ch. Tien, J. F. Bauters, M.J.R. Heck, D. T. Spencer, D. J. Blumenthal, and J.E. Bowers, ``Ultra-high quality factor planar $Si_3N_3$ ring resonators on Si substrates,'' Opt. Express 19 (2011) 13551--13556.



\bibitem{xu2} Q. Xu, D. Fattal, R.G. Beausoleil, ``Silicon microring resonators with \SI{1.5}{\micro\metre} radius,'' Opt. Express 17 (2008) 4309--4315.


\bibitem{gondarenko} A. Gondarenko, J. S. LEvy, and M. Lipson, ``High confinement micron-scale silicon nitride high Q ring resonator,'' Opt. Express 17 (2009) 11366--11370.



%
%

\bibitem{lewin} L. Lewin, D.C. Chang, E.F. Kuester, ``Electromagnetic waves and curved structures,'' Peter Peregrinus Ltd., Stevenage, England, 1977.

\bibitem{heiblum} M. Heiblum, Y.H. Harris, ``Analaysis of curved optical waveguides by conformal transformation,'' IEEE J. Quantum. Electron. QE-11 (1975) 75--83.

\bibitem{marcuse} D. Marcuse, ``Bending losses of the asymmetric slab waveguide,'' Bell Syst. Tech. J. 50 (1971) 2551-2563.


 
\bibitem{marcusec} D. Marcuse, ``Directional Couplers Made of Nonidentical Asymmetric Slabs.  Part I: Synchronous couplers,'' J. Lightwave Technol. LT-5 (1987) 113--118.



%




%
%


%
%
%

\bibitem{tamir} T. Tamir (Ed.), Guided-Wave Optoelectronics, Second Ed., Springer, Berlin, 1990.
%
%

\bibitem{meep} A. F. Oskooi, D. Roundy, M. Ibanescu, P. Bermel, J. D. Joannopoulos, and S. G. Johnson, ``MEEP: A flexible free-software package for electromagnetic simulations by the FDTD method,'' Computer Physics Communications 181 (2010) 687–702.

\bibitem{harminv} V. A. Mandelshtam and H. S. Taylor, ``Harmonic inversion of time signals,'' J. Chem. Phys. 107 (1997) 6756-6769.

\bibitem{xia} F. Xia, L. Sekaric, and Y.A. Vlasov, ``Mode conversion losses in silicon-on-insulator photonic wire based racetrack resonators,'' Opt. Express 14 (2006) 3872--3886.

\end{thebibliography}
\end{document}